\documentclass[%
 reprint,amsmath,amssymb,aps,superscriptaddress,nofootinbib]{revtex4-2}
\pdfoutput=1

\usepackage[utf8]{inputenc}
\usepackage[T1]{fontenc}

\usepackage{amsmath}
\usepackage{amssymb}
\usepackage{graphicx}
\usepackage{xcolor}
\usepackage{float}
\usepackage{multirow}
\usepackage{ulem}

\newcommand{\beq}{\begin{equation}}
\newcommand{\eeq}{\end{equation}}
\newcommand{\fr}{\frac}
\renewcommand{\d}[1]{\ensuremath{\operatorname{d}\!{#1}}}

\begin{document}

\title{Properties of dynamical fractal geometries in the model of Causal Dynamical Triangulations}

\author{J.~Ambjorn}
\email{ambjorn@nbi.dk}
\affiliation{{\small{The Niels Bohr Institute, Copenhagen University, Blegdamsvej 17, DK-2100 Copenhagen Ø, Denmark; IMAPP, Radboud University, Nijmegen, PO Box 9010, the Netherlands}}}
\author{Z. Drogosz}%
 \email{zbigniew.drogosz@doctoral.uj.edu.pl}
 \author{A. Görlich}%
 \email{andrzej.goerlich@uj.edu.pl}
 \author{J. Jurkiewicz}%
 \email{jerzy.jurkiewicz@uj.edu.pl}
\affiliation{%
{\small{Institute of Theoretical Physics, Jagiellonian University, \L ojasiewicza 11, Kraków, PL 30-348, Poland.}}}%

\date{\today}


\begin{abstract}
We investigate the geometry of a quantum universe with the 
topology of the four-torus. The study of non-contractible 
geodesic loops reveals that a typical quantum geometry consists
of a small semi-classical toroidal bulk part, dressed with 
many outgrowths, which contain most of the four-volume and which 
have almost spherical topologies, but nevertheless are quite 
fractal.
\end{abstract}

\maketitle

\section{Introduction}
Causal Dynamical Triangulations (CDT) is a model that 
attempts to apply methods of Quantum Field Theory
in the context of a quantum model of geometric degrees of 
freedom\footnote{Reviews of the model can be found in \cite{reviews1, reviews2}. The main idea is to have a lattice model
of quantum gravity where one in a non-perturbative way can 
test the idea of asymptotic safety \cite{weinberg, asymp-safe, reuteretc, reuter2007, Niedermaier:2006wt, Litim2004}.
Models of Dynamical triangulations (DT) were earlier attempts
in this direction \cite{aj, am, Agishtein:1992xx}, which however did not work, but 
see \cite{Ambjorn:2013eha, Coumbe:2014nea, Laiho:2016nlp} for recent attempts to revive that class
of lattice models.}.
As will be described below, the model comes with a (proper) time,
whereas the description of the geometries in the spatial directions is genuinely coordinate independent. The existence of the time coordinate has been instrumental for the construction of an effective mini-superspace action of the quantum theory, where we have integrated over the spatial geometries. In particular, it allowed us to talk about the emergence of a semi-classical mini-superspace geometry, as well as quantum fluctuations thereof \cite{Ambjorn:2001cv, Ambjorn:2005qt, Ambjorn:2004qm, agjl, Ambjorn:2007jv}. 
There is no reason not to expect  
a similar emergence of geometry in the spatial directions.
However, bearing in mind the importance of the proper time 
coordinate in our analysis of the mini-superspace geometry, it might be preferable to reintroduce some aspects of coordinates in the spatial directions, too. In some sense this is against the 
spirit of General Relativity which is coordinate independent,
but coordinates {\it can} be very useful.

In our recent paper \cite{coordinates} we discussed
one possibility of reintroducing coordinates in CDT in the spatial directions. In many of the former studies of four-dimensional CDT, the spatial topology was chosen to be 
that of a three-sphere $S^3$, and, given a spatial geometry as it appears in the path integral, we know of no simple way of reintroducing useful spatial coordinates in that case. However, in \cite{coordinates} the spatial 
topology of the Universe was chosen to be that of a three-torus $T^3$.
A $d$-dimensional manifold with a toroidal topology can be viewed as consisting of an elementary cell,
which is periodically repeated infinitely many times in all $d$ directions.
Although the choice of the elementary cell is not unique, the possibility of introducing
such an object enables the use of its boundaries as a reference frame,
with respect to which a Cartesian-like system of coordinates determined by the geodesic
distance to the boundaries may be constructed.
One conclusion drawn from the analysis in \cite{coordinates} was that the geometry of a typical 
triangulation which appears in the CDT path integral 
is surprisingly fractal. Before trying to extract any emergent spatial geometry from such triangulations it is thus important 
to understand the spatial geometry of a typical quantum configuration better. For that purpose we have found it advantageous
to use topological observables:
closed non-contractible geodesic loops, connecting the same geometric object in different copies of the elementary cell. 
The distribution of the length of shortest loops with a given set of winding numbers
passing through particular elements of geometry yields
information about geometric structures, and this kind of analysis has been used successfully in the study of two-dimensional 
Euclidean quantum gravity \cite{ab}.

\begin{figure}[]
\centering
\includegraphics[height=55mm]{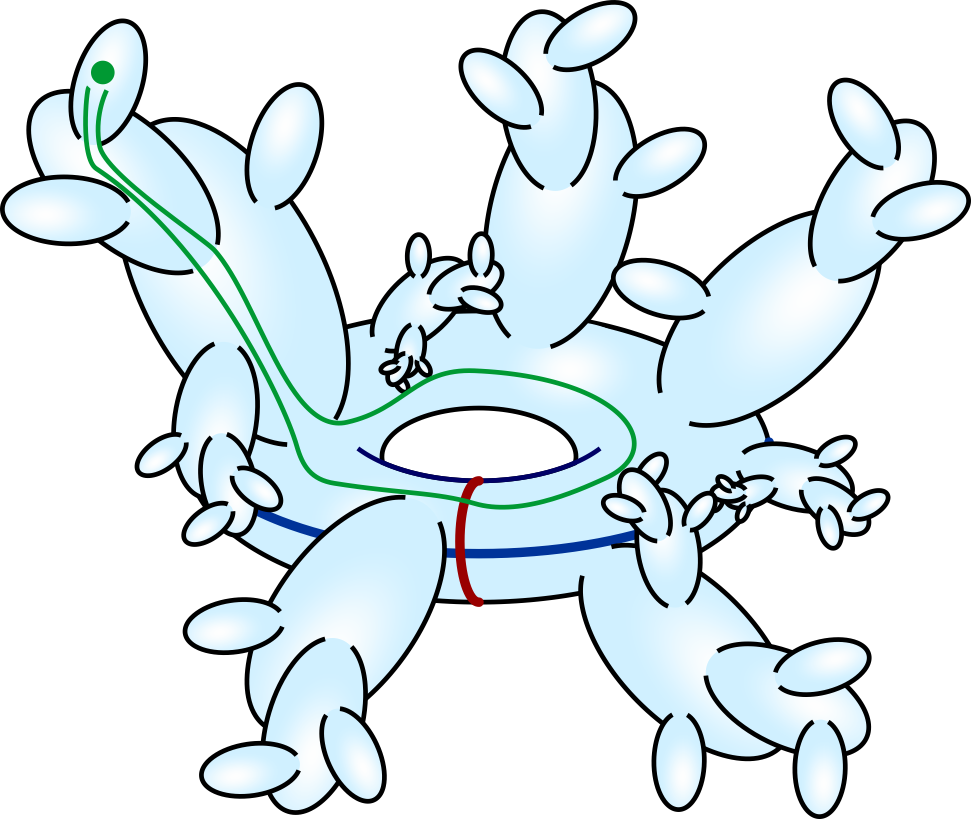}
\includegraphics[height=55mm]{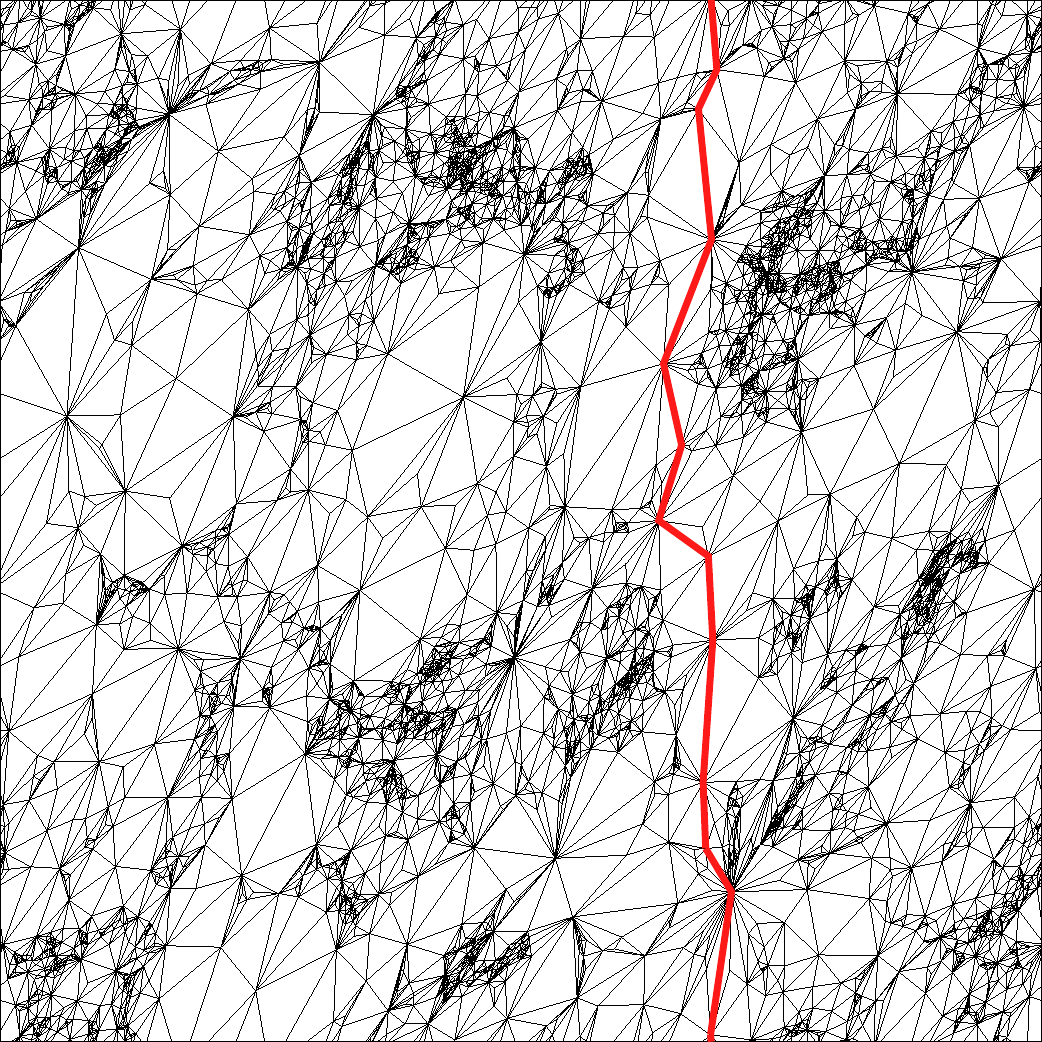}
\caption{
Up: illustration of a torus with outgrowths. 
The blue and red lines represent two non-equivalent and non-contractible loops.
The green loop is the shortest loop passing through the green point in the same direction as the blue line.
Down: embedding of a
triangulation of the two-torus consisting of 150000 triangles into the 
Euclidean plane (picture from \cite{ab}).
Shown in red is the shortest non-contractible loop.}
\label{fignew}
\end{figure}

The upper part of Fig.~\ref{fignew} provides a two-dimensional illustration of what we are looking for in the case of the higher dimensional tori.
We imagine that we have an underlying ``semi-classical'' toroidal structure, but there can be many outgrowths, which can be viewed as its quantum fluctuations.
A point in an outgrowth will have a long non-contractible geodesic loop passing through it,
while a point on the ``semi-classical'' toroidal part will have a short non-contractible
geodesic loop.
In this way one can map out the geometry of the 
toroidal universe in considerable detail, as will be described below. 
An example point in an outgrowth is marked with a green dot in the upper part of Fig.~\ref{fignew}.
and the green line is a non-contractible geodesic loop for this point.

In the lower part of Fig.~\ref{fignew} we have 
shown how a two-dimensional toroidal quantum configuration looks.
The configuration is a two-dimensional triangulation made of 
150000 equilateral triangles, generated
by Monte Carlo simulations of two-dimensional Euclidean quantum gravity. By a conformal mapping the triangulation 
can be mapped to an elementary cell in the plane. What is 
shown is a piecewise-linear approximation to this mapping (plus an affine mapping to make it a square).
The figure illustrates how such a quantum configuration consists of \textit{mountains} (outgrowths) and \textit{valleys}.
By far the most two-volume (the greatest number of triangles) is contained in the outgrowths, as can readily be seen from the picture.
In four-dimensional CDT,
we consider paths that connect centers of simplices, i.e.,
which consist of edges of the dual triangulation (see Sec.~\ref{model}).
In the two-dimensional case, the red line shown
in the lower part of Fig.~\ref{fignew}
consists of links of the direct lattice.
The picture shows quite precisely the fractal structure 
of two-dimensional quantum gravity. It is known that the Hausdorff dimension of spacetime in two-dimensional quantum gravity is {\it four} and not two, as one might perhaps naively expect \cite{hausdorff, Ambjorn:1995dg, Ambjorn:1995rg}. On a regular torus consisting 
of $N$ triangles one would expect a shortest loop of length 
approximately $N^{1/2}$ links. However, here we see that the
length is much closer to $N^{1/4}$. In particular, this implies
that the number of triangles in the valleys scale as $N^{1/2}$,
and not proportionally to $N$. The area of the valleys will thus 
disappear in an $N \to \infty$ limit where the continuum 
area $V \propto N a^2$ is kept fixed, $a$ being the length of 
a link in the triangulation before it was projected onto the plane. Therefore, in the two-dimensional case the valleys are
{\it not} semi-classical, but a quantum phenomenon.
We expect the situation to be different in the case of a
four-dimensional CDT torus, the reason being that the Hausdorff
dimension of a typical CDT configuration is four, i.e.\ 
the same as the canonical dimension of the spacetime. We {\it might} then have a picture where the valleys of $T^3$ constitute a semi-classical configuration which can act as a starting point for a description of a semi-classical spatial geometry. This is one 
of the points we will investigate in this article.

The rest of the article is organized as follows: in 
Sec.~\ref{model} we shortly define the CDT model of quantum 
gravity, in order to fix the notation (we refer to \cite{reviews1, reviews2} for more detailed definitions). In Sec.~\ref{topology} we define
certain characteristics which are special for spacetimes with toroidal topologies. Sec.~\ref{sec:simulations}  describes how 
the Monte Carlo simulations are performed, whereas Sec.\ 
\ref{shortestloops} reports on the measurements of the shortest loops of winding number one. In Sec.~\ref{sec:higher} these
measurements are generalized to loops with higher winding numbers. In Sec.~\ref{alternativeBC} we generalize even the 
possibilities of higher winding numbers, acknowledging the fact that our winding numbers are depending of our chosen reference frame and that a true geometric winding can be any linear combination of our labeling of windings. Sec.~\ref{neighborhood} discusses if simplices in the outgrowths and simplices in the valleys 
have different geometric neighborhood, possibly signifying that 
the valleys can be viewed as semi-classical, while the outgrowths might be viewed as quantum fluctuations. Finally, Sec.~\ref{conclusion} contains a discussion of the results and our conclusions.

\section{The model}\label{model}

The basic idea in CDT is to calculate the quantum amplitude of the transition between two physical states. 
The amplitude is defined as a path integral over field configurations, which in this case
are spacetime geometries,
\beq
\mathcal{Z} = \int \mathcal{D}[g_{\mu \nu}] e^{iS_{EH}[g_{\mu \nu}]}.
\label{eq:z}
\eeq
$S_{EH}$ is the Einstein-Hilbert action
\beq
S_{EH} [g_{\mu \nu}] = \fr {1}{16 \pi G} \int_M \d{}^4 x \sqrt{- \det \ g} (R-2 \Lambda ),
\eeq
where $R$ is the scalar curvature and $\Lambda$ is the cosmological constant.  This expression is formal and requires regularization and a precise definition of both the integration measure
over $g_{\mu\nu}$ and the domain of integration over spacetimes.  In CDT it is assumed that we will take into
account only spacetimes that admit a global time 
foliation: $M = \Sigma \times I$.
The term {\it causality}
in the context of the model means that the topology of space $\Sigma$ is preserved in time evolution.
An additional assumption is that the spatial topology of the Universe is closed.
Corresponding to $I$ there is an initial
and a final global time for the geometries considered, and the
amplitude (\ref{eq:z}) is the transition amplitude between
the spatial geometries at the initial and final global times.
This amplitude can be calculated analytically if spacetime is 
two-dimensional \cite{al}, but in the case of three- or 
four-dimensional spacetime we have to rely on numerical simulations, and thus a discretization of spacetime geometries.

The spacetime geometries are discretized using a method based on an idea of Regge  \cite{Regge},  and the diffeomorphism-invariant integral over metrics 
(\ref{eq:z})  is regularized by a sum over a set of simplicial manifolds with a correct topological structure.
For each spacetime of this kind it is
possible to perform Wick rotation to Euclidean signature, after which
the exponent in the sum becomes real and the complex amplitudes become real probabilities (see \cite{ajl1} for details):
\begin{equation}
{\cal{P}}({\cal T}) \propto e^{-S({\cal T})}. 
\label{eq:prob}
\end{equation}
This formulation is well-suited to numerical 
simulations, which, as mentioned, are the 
main tool used in the analysis. 
The foliation of spacetime 
defines an ordering on the slices (leaves) $\Sigma$, each of which can in a natural way be assigned an integer time 
parameter $t$.

In the 3+1-dimensional case, the spacetime is built out of four-dimensional simplices.
Each of them is the convex hull of five vertices that lie on two neighboring slices $\Sigma$.
There are thus two types of four-simplices: $\{4,1\}$-simplices with four vertices on a slice $t$
and one vertex on a slice $t\pm 1$, and $\{3,2\}$-simplices with three vertices on a slice $t$
and two vertices on a slice $t\pm 1$. 
Each simplex abuts along its 
three-dimensional faces on five other simplices, called its neighbors.
All space-like links, i.e., line segments which connect two vertices
on the same time slice, are of length $a_s$, and all time-like links, i.e., line segments
which connect two vertices on neighboring time slices, are of length $a_t$.
Those lengths are unchanging, and their ratio squared is the asymmetry factor:
$\alpha = a_t^2/ a_s^2$.

The Regge action (the Hilbert-Einstein action on a piecewise linear manifold)
for a causal triangulation depends only on global quantities:

\begin{equation}
\begin{aligned}
S_{EH}({\cal T}) = & -(K_0+6\Delta) N_0 + K_4 \left( N^{\{4,1\}}+ N^{\{3,2\}}\right) \\
& +\Delta \cdot  N^{\{4,1\}},
\end{aligned}
\label{eq:seh}
\end{equation}
where $N_0$, $N^{\{4,1\}}$ and $N^{\{3,2\}}$ denote the total number of vertices and of
$\{4,1\}$- and $\{3,2\}$-simplices in the configuration.
The three dimensionless coupling constants, $K_0$, $K_4$ and $\Delta$, are related respectively to 
the inverse of the gravitational constant $G^{-1}$, the cosmological constant $\Lambda$, 
and the asymmetry factor $\alpha$.

To describe a configuration fully, one has to do the following:
\begin{itemize}
    \item choose the initial and final states. To avoid the problem of making such a choice, we customarily adopt the periodic boundary conditions with some number of time slices $T$;
    \item label all the vertices and all the four-simplices;
    \item list all the vertex labels together with corresponding time parameters;
    \item list all the four-simplex labels together with the quintuples of 
    their vertices
    and their neighbors placed opposite to the vertices.
\end{itemize}
The same data are contained in the dual description, which is a graph (called the dual lattice) whose 
vertices correspond to the four-simplices of the configuration, 
and whose links correspond to interfaces between the four-simplices. 

As mentioned, no analytic solution for the model exists in 3+1 dimensions. Therefore, we probe 
the trajectory space by random generation of configurations with desired topology
and scrutinize the results. 
The configurations are not created one-by-one from scratch, but instead they
are generated in large number 
by performing a Monte Carlo simulation, which starts from a very simple triangulation and
lets it gradually evolve by means of 7 types of geometric moves.
The moves modify the configuration locally
in a topology-preserving way and are ergodic, which means that by performing them it is possible
to obtain any triangulation with the same topology. 
In every simulation many billions moves are performed, which allows us to overcome 
auto-correlation and to generate independent configurations.
Moves are performed at random in a way satisfying the detailed balance condition and with correct probabilities derived from the action. We 
set the values of the couplings $K_0$ and $\Delta$
before starting the simulation in order to study the model at a chosen point in the coupling constant space
(cf. \cite{phase}).
The number of triangulations grows exponentially with 
$N_4=N^{\{4,1\}}+ N^{\{3,2\}}$
for a fixed topology\footnote{There is no analytical proof of this, only numerical evidence \cite{aj94}.}.
Summing over all triangulations with a fixed $N_4$, using as 
a weight $e^{-S_{EH}( {\cal T})}$ for each triangulation
${\cal T}$, will
result (to leading order in $N_4$) in an expression
\begin{equation}
{\cal Z}_{N_4}(K_0,\Delta,K_4)\propto e^{(K_4^{crit}(K_0,\Delta)-K_4)N_4},
\label{exp}
\end{equation}
and the full discretized version of (\ref{eq:z}) is then 
\begin{equation}
{\cal Z}(K_0,\Delta,K_4) = \sum_{N_4}
{\cal Z}_{N_4}(K_0,\Delta,K_4).
\end{equation}
In general we are interested in the limit where the average value of $N_4\to \infty$, which corresponds to the limit
$K_4 \to K_4^{{crit}^+}$. In simulations, taking this limit is replaced by studying the properties of a sequence of spacetimes, each with a fixed $N_4$ and $N_4\to \infty$.

\section{Topology, boundaries and coordinates}\label{topology}

In the original formulation of CDT in four dimensions it was assumed that the spatial topology of time slices was spherical ($S^3$).
For technical reasons, related to the computer
simulations, it was assumed that time was periodic. However, this periodicity played no role in the initial study of universes 
with $S^3$ topology as long as the time period was sufficiently large.
The existence of the time foliation sufficed to analyze the phase structure of the model.
The phase diagram is surprisingly complex when one considers the extreme simplicity of the action (\ref{eq:seh}),
which only depends on the global quantities  $N_0$, $N^{\{4,1\}}$ and $N^{\{3,1\}}$.
Of the four different phases, only one, the so-called $C$ phase (also called the de Sitter phase) seems to be related in a simple way to semi-classical spacetimes,
and we will here only discuss results obtained when the coupling constants are chosen such that the system is in this phase.
The simplest observable measured was the spatial volume profile $N_3(t)$, defined as the number
of spatial tetrahedra on a time slice $t$. 
A typical system with a sufficiently large number of slices $T$
consisted of a {\it blob} and a cut-off size {\it stalk} (necessary to satisfy the periodic time boundary 
conditions mentioned above). Owing to the invariance with respect to (discrete) translations in time, the position
of the blob could be arbitrarily shifted in time. We used this possibility to center 
it around a fixed time position. It was shown that both the average volume
$\langle N_3(t)\rangle$ and its fluctuations can be derived from the discretized version of the
effective mini-superspace action \cite{mini} for the isotropically homogeneous 4D universe. The classical solution in this
case corresponded to a 4D de Sitter sphere.
It should be noted that although the mini-superspace model was originally proposed in the context of 4D General Relativity
\cite{hartle},
where all degrees of freedom except the scale factor were suppressed, the situation in CDT is different.
The first difference is that $N_3(t)$ corresponds to a collective state, where all degrees of freedom are integrated out. The second difference is the sign of the effective action, opposite to 
the one found
in General Relativity. In CDT the solution of classical equations of motion gives a 
stable classical vacuum state, where at each $t$ we have all possible geometric realizations
with a particular value of $N_3(t)$. The existence of this highly nontrivial classical General 
Relativity limit of the model was one of the most important results in the early studies of 
CDT.

It is an interesting question whether the semi-classical limit can be extended to
include degrees of freedom in spatial directions. The simplicity of the $S^3$ topology, however,
makes such analysis very difficult or even impossible to perform
because of the background
independence. We do not have any reference system with respect to which observables
could be measured. This may be different if we decide to formulate the model with a
richer spatial topology.
In the analysis presented in this article
we chose $\Sigma = T^3$, and for technical reasons (ease of computer implementation
and eschewal of the need for selecting the initial and the final states) we imposed periodic boundary conditions in the time direction: $\mathcal{M} = T^3 \times T^1$.

Thus, each configuration is topologically a Cartesian product of four circles.
Each closed curve within the configuration is homotopically equivalent to
a combination of those circles, and the coefficients of that linear combination are the four winding numbers of the loop.
Let us call them the winding numbers in the $x$, $y$, $z$ and $t$ directions.

One can equivalently consider the four-torus as an infinite periodic system.
All the $N_4$ simplices of the torus are contained in an elementary cell, 
which repeats itself infinitely many times in four directions.
The elementary cell can be defined in various ways, each of which is 
equivalent to a choice of a set of faces between neighboring 
four-simplices to form the cell's four boundaries. 
A loop within the torus corresponds in this picture to a path 
joining the same simplices in two different copies of the elementary cell.
We can assign a set of four numbers to each copy of the elementary cell
in such a way that the differences between them for any two copies 
are identical to the four winding numbers of the corresponding loop. 

Arguably it is the most convenient to look at loops in the dual picture, and so
henceforth we will most often use the word ``loop'' to mean not a 
spacetime curve but an ordered set of connected simplices whose image in the dual lattice is a non-contractible directed cycle.
The length of a loop is the number of links in the cycle. (For simplicity we assume that all links
have the same length.)
Similarly, a geodesic between two simplices will mean a line connecting them whose 
image in the dual lattice has minimal length.

\begin{figure}[]
{\includegraphics[width= 0.95 \columnwidth]{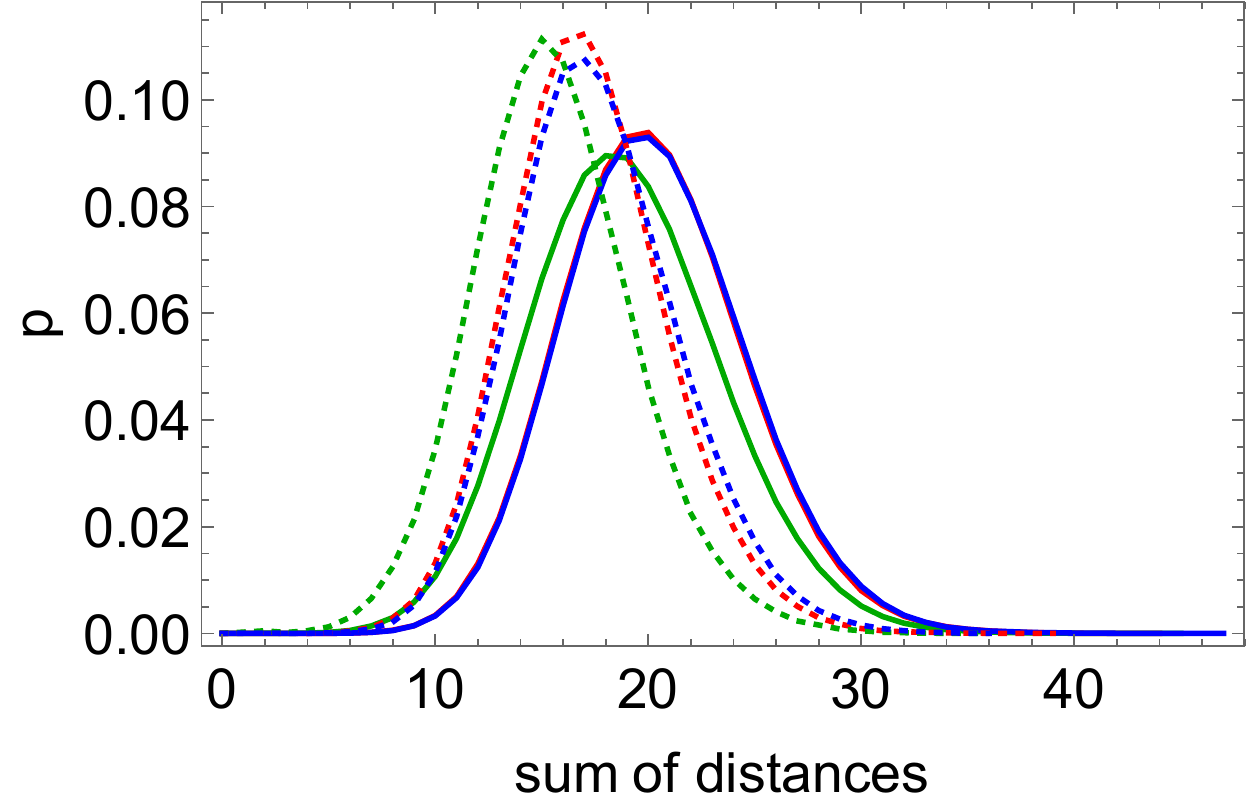}}
{\includegraphics[width= 0.95 \columnwidth]{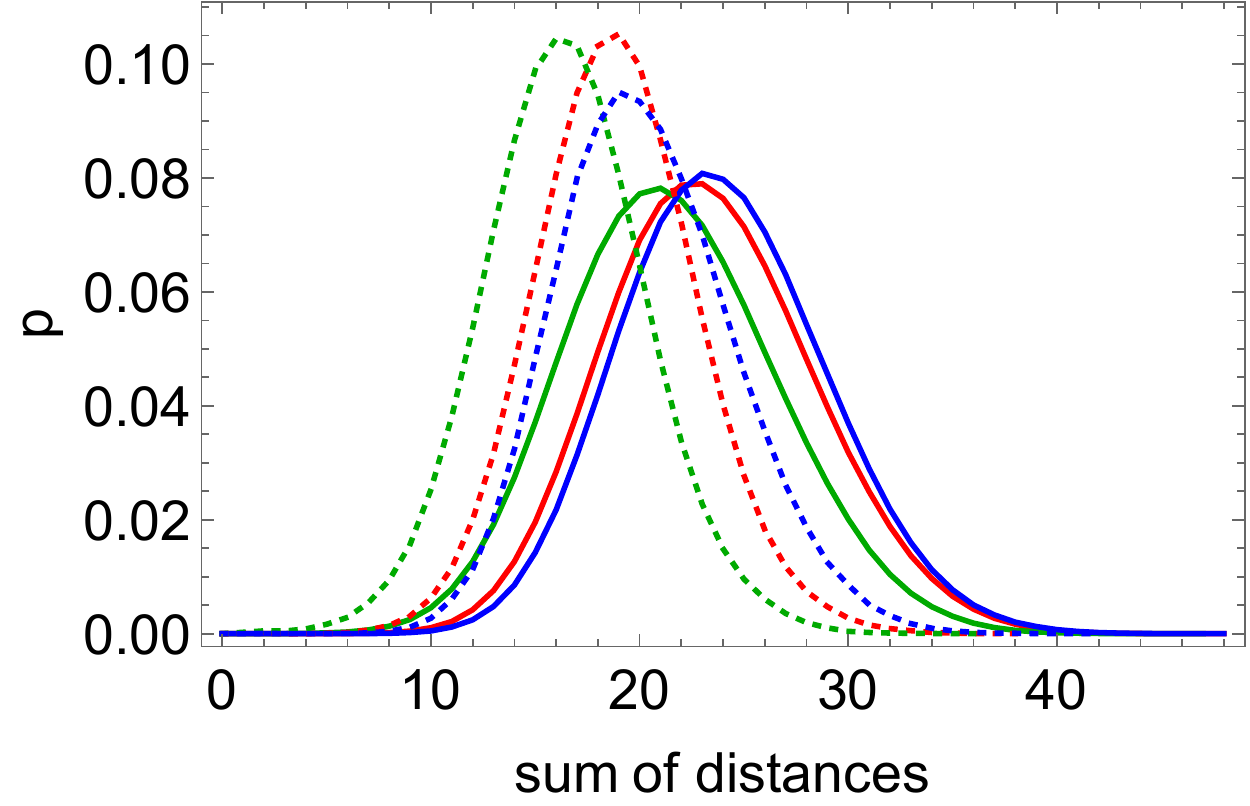}} 
\caption{Distributions of a sum of distances from simplices to the two opposite boundaries
in the $x$ direction (red), the $y$ direction (green) and $z$ direction
(blue) for systems with $N^{\{4,1\}}=80\mathrm{k}$ (up) and $N^{\{4,1\}}=160\mathrm{k}$ (down).
The distributions scale consistently with the Hausdorff dimension $d_H=4$.
The dotted lines refer to simplices adjacent to the boundary,
$x = 1$ or $x' = 1$ respectively for the two sides of the boundary.}
\label{fig:shifts}
\end{figure}

The distance (the minimal number of links in the dual lattice) from a simplex to each of the
boundaries of the elementary cell serves as pseudo-Cartesian coordinates of the simplex. 
This definition was studied in a previous paper \cite{coordinates}. In a regular hypercubic lattice
a sum of distances from any simplex to the two opposite boundaries is a constant,
equal to the geodesic distance between the boundaries.
On a random lattice generated by CDT this is, however, not the case.
We observe a nontrivial distribution of these values (see Fig.~\ref{fig:shifts}),
which may indicate either that the shape of the elementary cell is far from being rectangular,
or that quantum fluctuations of the geometry can be viewed as ''mountains`` and in effect simplices close to the top of the mountains have a larger geodesic
distance to the boundaries than those lying in the valleys between the mountains.
Results indicate that both effects may be important.
The latter effect is supported by the difference visible in Fig.~\ref{fig:shifts} between
the distributions $P(x + x')$ for all simplices (solid lines) and simplices adjacent to a boundary (dotted line, $x = 1$ or $x' = 1$).
Boundaries are chosen to locally minimize their area,
thus they \textit{prefer} the central region of a torus (\textit{valleys}) and omit outgrowths (\textit{mountains}).
Therefore, simplices adjacent to one boundary are closer to the second boundary 
than an average simplex.

In this paper we will try to perform a closer analysis of relevant structures produced in simulations
to understand the properties of the {\it quantum landscape}. Of a primary interest will be correlations between
valleys, which in this picture can be interpreted as a semi-classical background geometry.

\section{Description of simulations}\label{sec:simulations}

\begin{figure}[]
\centering
\includegraphics[scale=0.5]{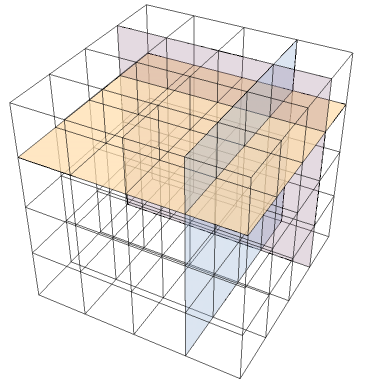}
\caption{A schematic view of a single time slice in the initial configuration. Visible are the $4^3$ hypercubes (each of which is divided into 16 simplices) and the starting position of the boundaries.}\label{fig:cube}
\end{figure}

The starting point of all the Monte-Carlo simulations was a single
simple configuration, which contained 4096 simplices in
$4^4$ regularly placed four-dimensional hypercubes.
The considered configuration consisted of $T = 4$ time slices.
Interfaces between some neighboring simplices were chosen as the boundaries
of the elementary cell (cf. Fig.~\ref{fig:cube}). The precise shape of the initial configuration
and the initial position of the boundaries can be chosen freely
as long as they have the correct topology.

The boundaries are encoded as an additional set of numbers assigned to every dual link
in a triangulation.
Each link $\{ij\}$ in a dual
lattice is characterized by a set of four numbers $n_{ij}^\mu = \pm 1,0$, where nonzero values mean that
the link crosses the corresponding boundary in a positive or negative direction.
Here, $\mu = 1, \dots, 4$ enumerates directions.
These numbers
have an obvious property $n_{ij}^\mu=-n_{ji}^\mu$, and their sums along a closed loop reproduce
the winding numbers of the loop.

In order to keep the size of the boundary small, 
after each performed move
a procedure that changes the position of boundary
if more than two
faces of a single simplex belong to it was invoked
in the region affected by the move.

The simulations were performed at the canonical point in the phase space
of toroidal CDT, i.e., in the $C$ phase, for the parameters $K_0 = 2.2$ 
and $\Delta = 0.6$ and for $N^{\{4,1\}} = 160000$. For the analysis we chose a typical, well-thermalized configuration.
The total number of simplices of the configuration
we analyzed was equal to $N_4 =N^{\{4,1\}}+N^{\{3,2\}} = 370724$.

\section{Shortest loops}\label{shortestloops}

In a previous article \cite{coordinates} we introduced the idea of analyzing the shortest loops
of non-zero winding numbers passing
through a given simplex to gain understanding of the shape of the system. 
We described the distribution of lengths of loops with low winding numbers and 
noted the universality
of its shape. We also noted the strong correlations of distribution of loop lengths in different directions.
\begin{figure}[]
\begin{center}
{\includegraphics[width= 0.95 \columnwidth]{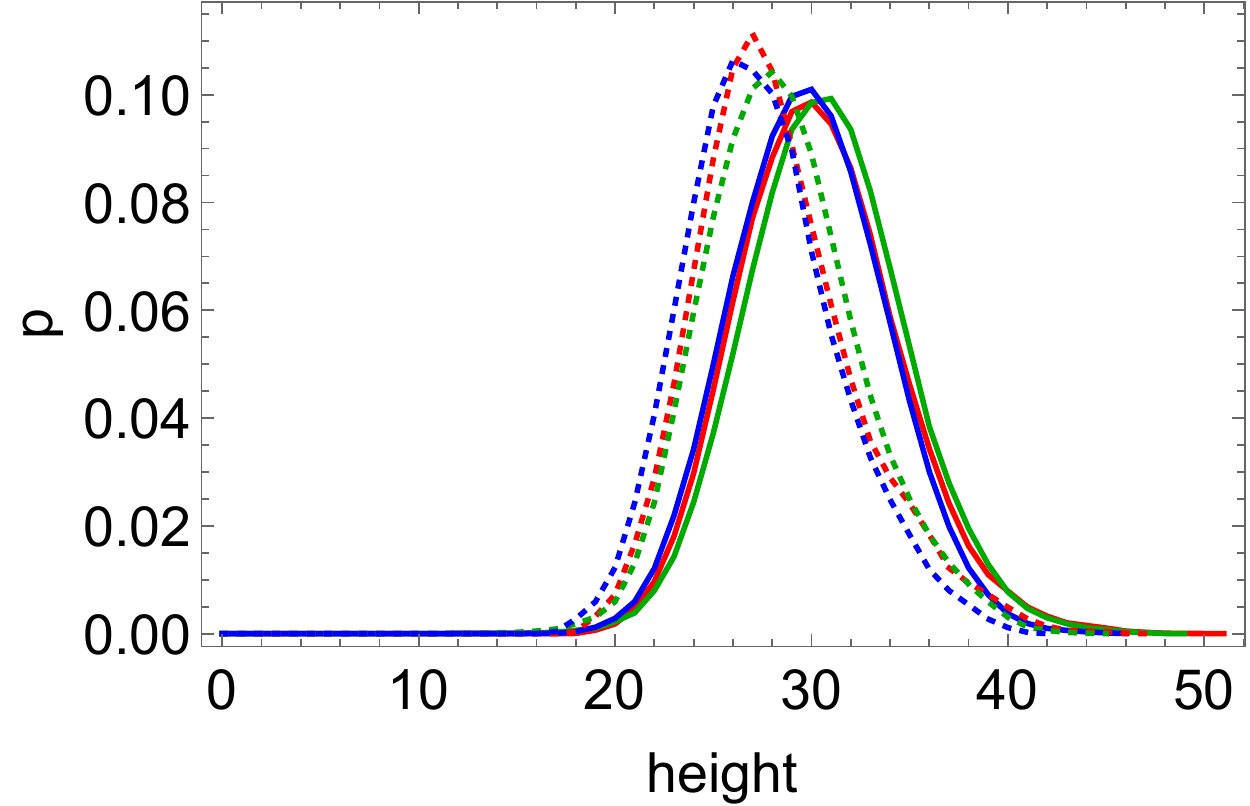}} 
{\includegraphics[width= 0.95 \columnwidth]{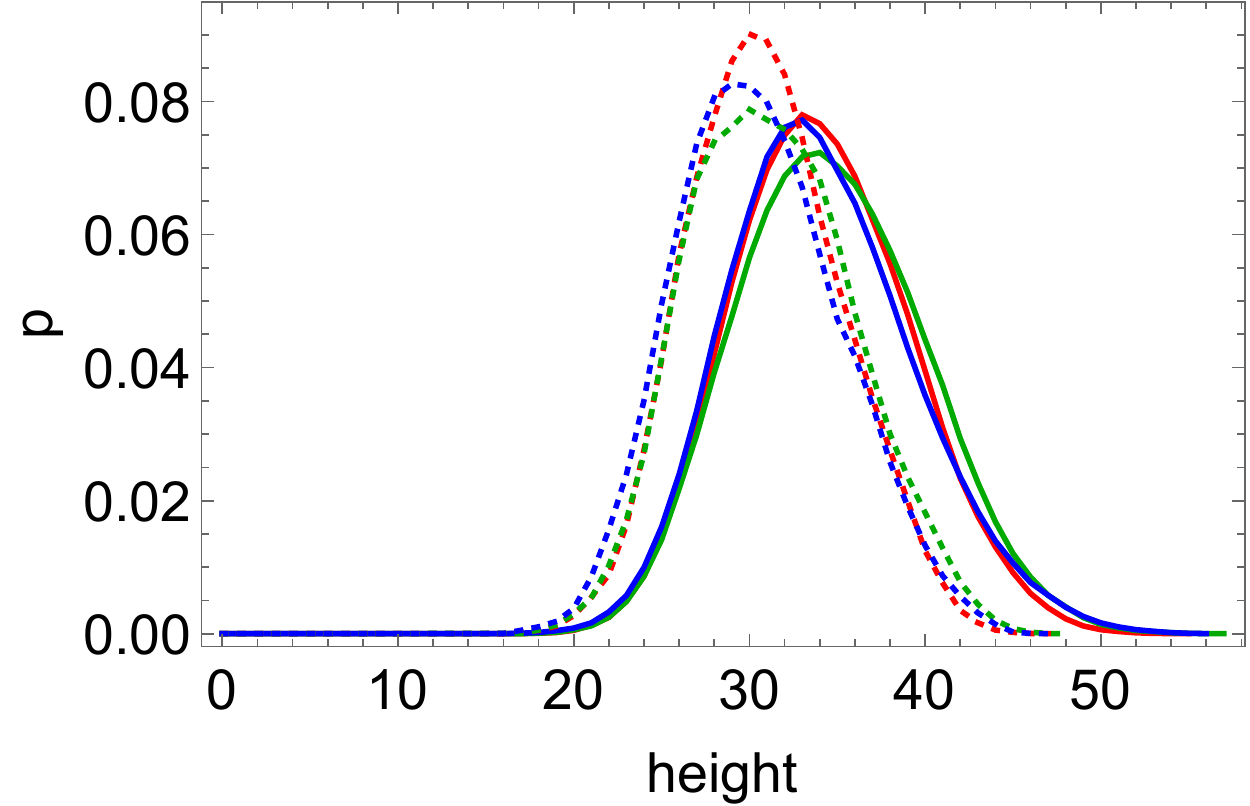}} 
\end{center}
\caption{Distributions of distances from simplices to their copies in neighboring elementary cells (heights) for systems with $N^{\{4,1\}}=80\mathrm{k}$ (up)  and $N^{\{4,1\}}=160\mathrm{k}$
(down). These are lengths of minimal loops with winding numbers $\{1,~0,~0,~0\}$ (red), $\{0,~1,~0,~0\}$ (green) and $\{0,~0,~1,~0\}$ (blue) shifted in $r$ by a shift of order one.
The dotted lines refer to simplices adjacent to the boundary.}
\label{wind}
\end{figure}

To recapitulate, in order to find the shortest loop of a given winding number
passing through a simplex, we treat the four-torus
as an infinite periodic system and follow step by step the front of a diffusion wave beginning at the chosen simplex
(using a diffusion wave in a system infinite in four directions is applicable to the case of low winding numbers;
otherwise this method becomes computationally inefficient and should be modified, cf. Sec.~\ref{sec:higher}). 
The number of loops with a given winding number that pass through a simplex grows (eventually) exponentially fast with the loop length.
Thus, while it is feasible to list all the shortest loops of a given type,
in the case of longer loops we usually have to pick one sample loop, representing their
universal properties.

\begin{figure}[]
\includegraphics[width=\columnwidth]{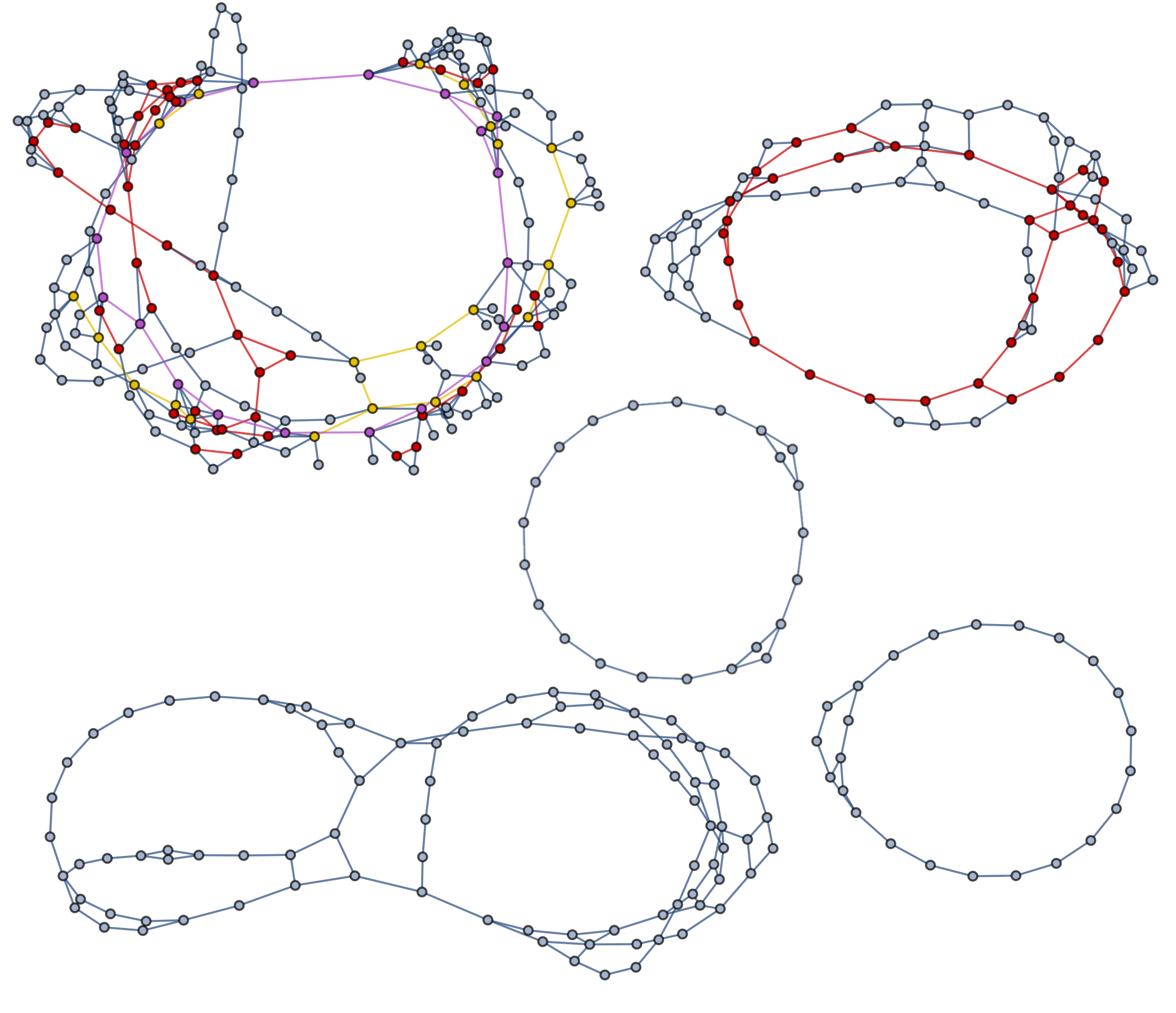}
\caption{The dual-lattice graphs of all the connections between simplices of lowest x-heights: 18 (purple), 19 (red), 20 (yellow) and 21 (blue). In general, loops longer than the minimal length almost always
contain fragments belonging to shorter loops, but for small heights there
exist also a small number of loops built only of simplices with equal x-height. }
\label{fig:shortest}
\end{figure}

Fig.~\ref{fig:shortest} presents the connections between simplices forming the shortest loops
of winding numbers \{1,0,0,0\} in the configuration.
It is evident that such very short loops are rare: in a configuration containing $N_4 = 370724$ simplices there are only 20 simplices belonging to loops of length 18. Moreover, loops of length
from 19 to 21 often differ from each other only by a few simplices;
the number of separate short loops of the same length – the number of distinct deepest
{\it valleys} – is very small.
The results in the other spatial directions are similar.

Geometry of the random manifold generated by computer simulation is highly fractal. 
It is tempting to
interpret the distribution of the lengths of loops with a unit winding number in spatial or time directions, having its maximum at a length above 30,
as a signal that in most cases the starting simplex is located inside one of the fractal structures
(mountains, outgrowths),
whereas the (rarer) simplices belonging to the loops with a length 
that is minimal or close to minimal
correspond to the (relatively simple) basic structure of valleys in the configuration. With this 
interpretation, the length of a loop starting from a particular simplex reflects the 
position of the simplex relative to the 
valleys. And so, for a particular configuration we assign to each simplex a unique set of four numbers: lengths of loops 
with a unit winding number in a particular direction and zero in the other three directions. Following these numbers along simplices belonging to any particular loop, we may
see how far the simplices belonging
to the loop are from the base, and how this distance changes along the loop. The four numbers
assigned to a simplex can be called its \textit{heights},
as they reflect its position above the basic structure. For the sake of brevity, we can use the names x-height, y-height,
z-height, t-height for the length of loops in the unit directions.
It was checked that, as expected, the height values of the five neighbors of 
any simplex differ from its own height by $\pm 2$, $\pm 1$ or 0.
In general any loop starting deep inside
a fractal structure is expected to move closer to the base and then eventually climb back to simplices in the same fractal. 
There are only few loops whose simplices are all of equal height. This property is possessed 
by the shortest loop in the configuration and a few dozen other short loops,
which are, so to speak, the ``locally deepest valleys''.
To summarize,
the \textit{height} of a simplex is defined as the length of the shortest loop passing through it.
It is determined separately for each topological dimension of the torus.
For a given simplex and direction, there might exist, and often do, several shortest loops,
all with the length equal to the height of the simplex.
In further analysis we pick only one of them for each simplex.
Usually, through each simplex pass also many loops (of the same length or longer) that are minimal for other simplices (see the discussion near the end of Sec.~\ref{alternativeBC}).

\section{Loops with higher winding numbers}\label{sec:higher}

The short loops contain important information about the underlying structure of the manifold and about the
distribution of valleys. The choice of four directions ($x,~y,~z$ and $t$) is 
nevertheless to a large extent arbitrary. It reflects
a particular structure of the initial configuration and the {\it memory} about the initially chosen elementary cell.
It may give a false impression that the elementary cell remains geometrically
hypercubic during the thermalization.
We can extend the analysis of minimal loop length distributions to include simplices in cells with an arbitrary set of winding numbers $\{n^\mu\}$. The analysis shows that the network of minimal loops contains not only loops in the four basic
directions, as discussed above, but also loops with nontrivial winding numbers. 
It is important to note, that the winding numbers of a loop
does not depend on a particular choice of boundaries or, equivalently, the elementary cell.

\begin{figure}[]
\centering
\includegraphics[width= 0.95 \columnwidth]{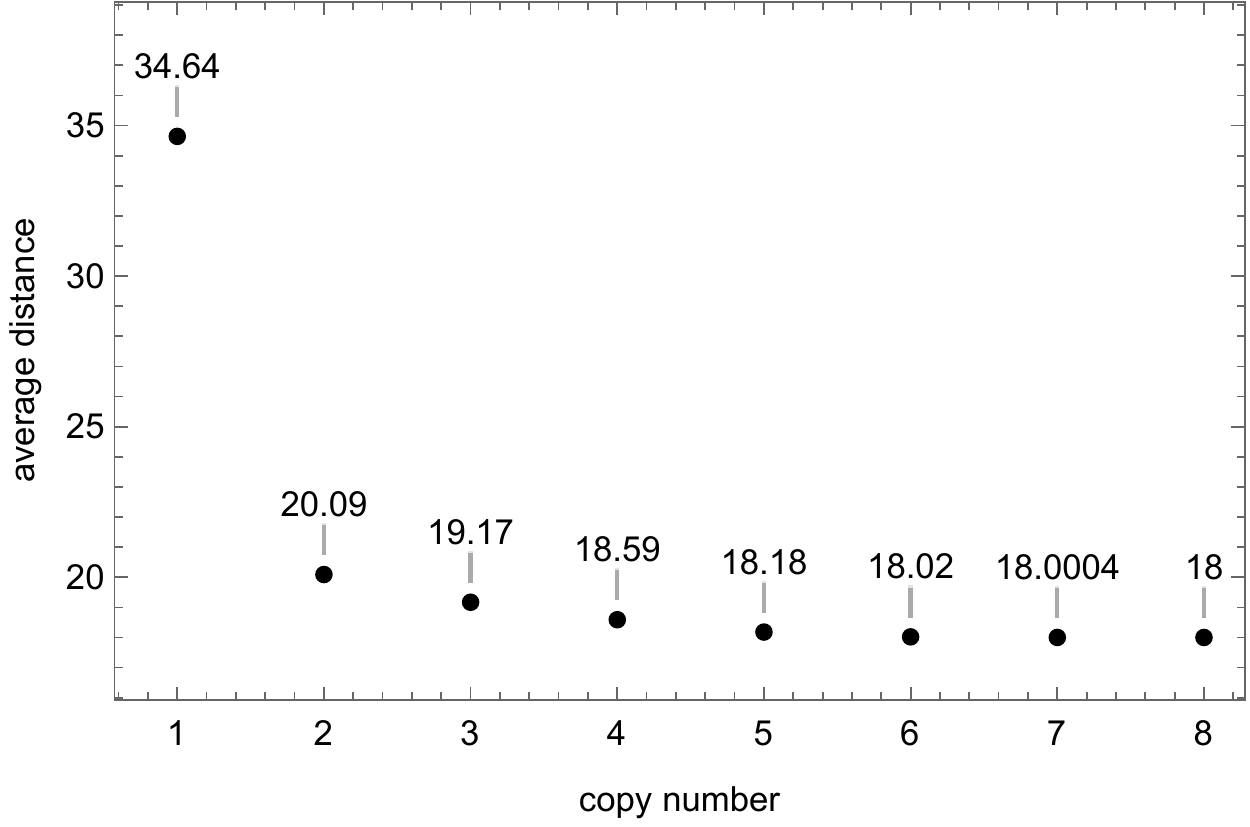}
\caption{Distance from a starting simplex to its copy in cell $\{n,0,0,0\}$ minus distance from the same simplex to its copy in cell $\{n-1,0,0,0\}$, averaged over all the simplices of the configuration.
Already for $n=8$ the minimal value of 18 is reached.}
\label{fig:av}
\end{figure}

\begin{figure}[]

\includegraphics[width= 0.95 \columnwidth]{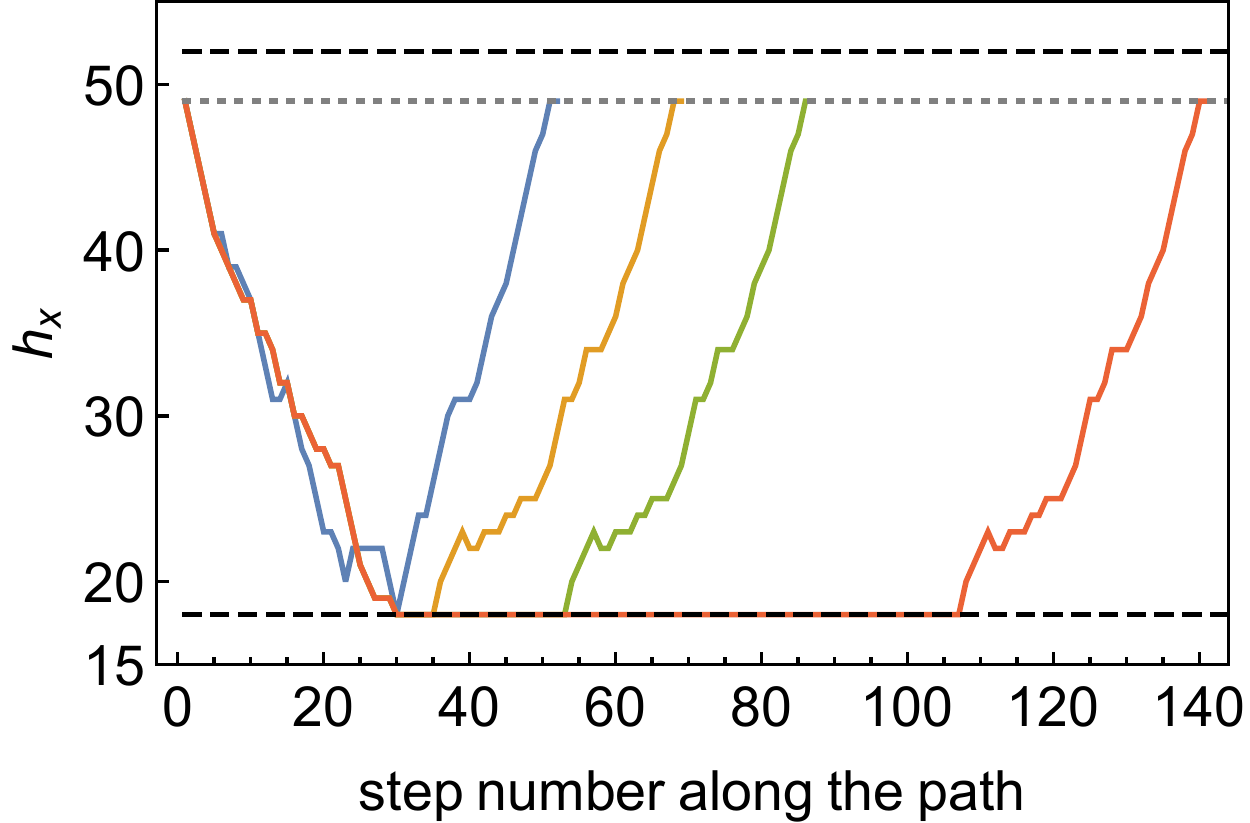} 
\includegraphics[width= 0.95 \columnwidth]{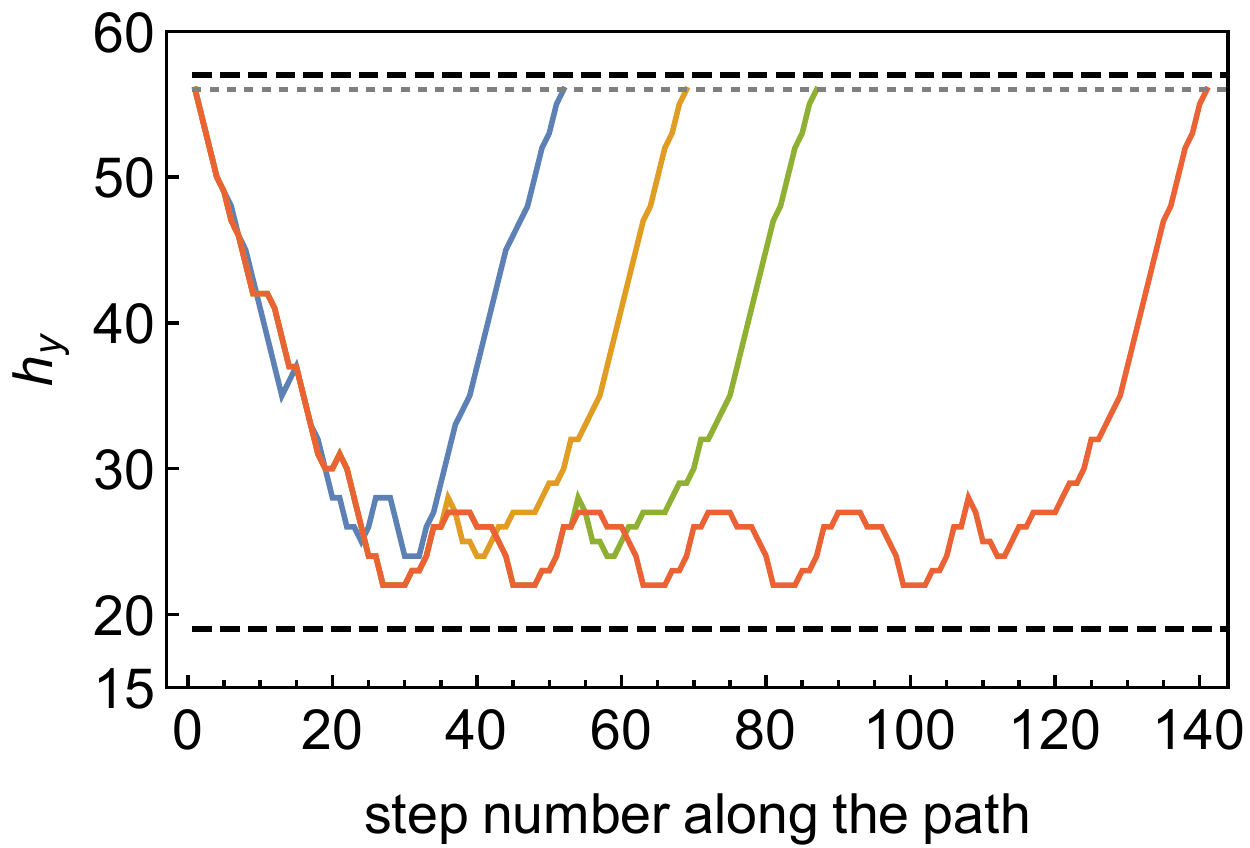}
\includegraphics[width= 0.95 \columnwidth]{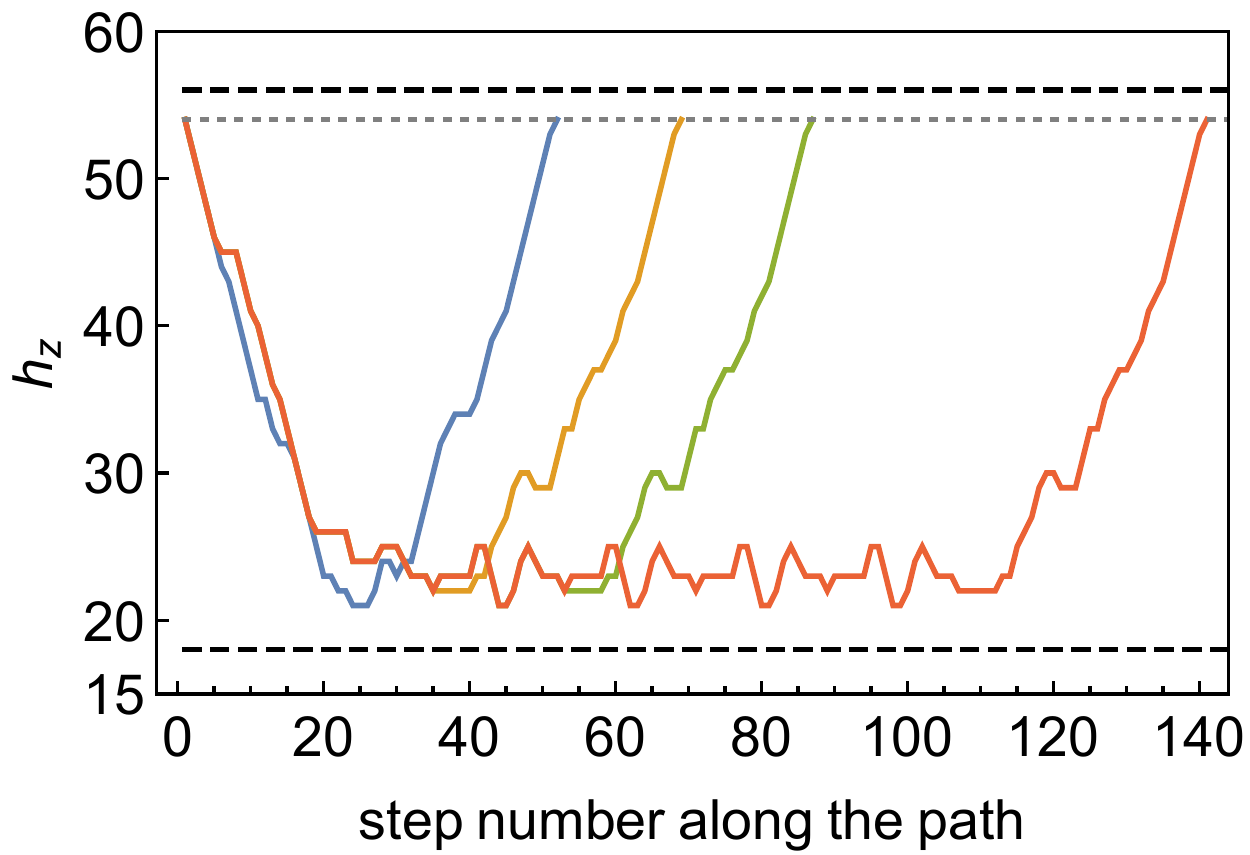} 
\includegraphics[width= 0.95 \columnwidth]{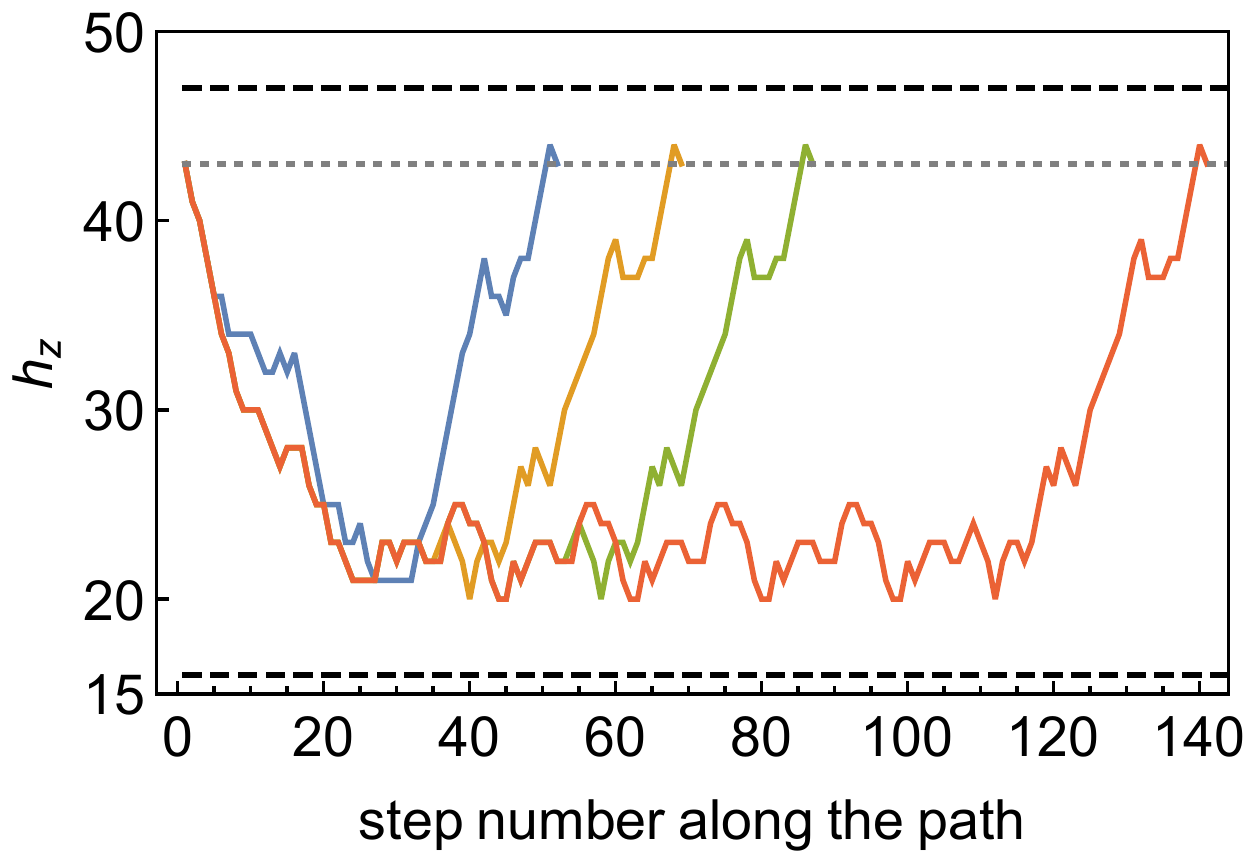}

\caption{Heights in the 4 basic directions of consecutive simplices along loops of winding
numbers $\{1,0,0,0\}$ (blue), $\{2,0,0,0\}$ (orange), $\{3,0,0,0\}$ (green), $\{6,0,0,0\}$ (red) starting from a simplex
in an outgrowth. The dashed lines indicate the minimal and maximal heights in the configuration,
and the dotted line indicates the height of the initial simplex.}
\label{fig:sim2}
\end{figure}

Using a four-dimensional diffusion wave in a system treated as infinite in four directions 
is a simple method to ensure that we find the shortest loop of a given winding number, regardless
of the shape of the dual lattice, as the diffusion wave cannot ``miss'' any short path. 
One could continue the diffusion to find loops with any higher winding numbers.
However, the number of visited simplices at a distance $R$ from the initial simplex grows as $R^3$,
which means that eventually, for large $R$, the procedure would become too time- and memory-consuming,
and computationally inefficient. We should, therefore, modify the boundary conditions
in such a way that the number of simplices visited by the diffusion wave grows more slowly.
One example is to consider the four-torus as a system infinite in only one direction –
for example the x-direction –
and strictly periodic in all other directions. This way we can measure loops with winding numbers 
$\{n,0,0,0\}$, $n=1,2,\dots$. A small modification of this idea is to assume that 
cell boundaries in all directions except $x$ are impenetrable for 
the diffusion wave.
In both of the methods the number of visited simplices 
in the $R^{\mathrm{th}}$ shell, for $R$ large enough, stabilizes and becomes independent of the distance $R$.
We will observe a faster growth only up to the range where the 
diffusion wave reaches the boundaries of the system in the finite directions. 
A similar result can be obtained if we put the impenetrable walls
in all directions except $x$ at boundaries between cells number $\pm 1$ and $\pm 2$.
This choice improves the behavior of the diffusion process in case the initial simplex is
near the boundary of the elementary cell.

In Fig.~\ref{fig:av} we present results of using this method (with impenetrable walls
in all directions except $x$ at boundaries between cells number $\pm 1$ and $\pm 2$) to measure the average lengths of loops with 
winding numbers $\{n,0,0,0\}$, $n=1,2,\dots,8$. Similar measurements were performed in
y- and z-directions.
It turns out that the difference between the distance to the copy number $n$ and the distance 
to the copy number $n-1$ of a given simplex decreases rapidly with increasing $n$, down to
the length of the minimal shortest loop in the given direction (lowest height of a simplex), and then it remains constant.
The explanation is simple: in order to minimize the length of a loop with a high winding number $n$ in a given direction,
it becomes advantageous for the loop to connect the initial simplex to a simplex of the 
lowest possible height,
then trace the shortest loop of unit winding number $n$ times, and finally return to the initial simplex.
As can be seen in Fig.~\ref{fig:av}, already for $n=8$ the shortest loop of the configuration
is a part of all the loops of winding number $n$. The figure shows the 
distances averaged over
all simplices in the configuration. We note that the average for loops with a unit winding number in the x direction for
this particular configuration is 34.64, which is well above the minimal length of 18,
showing that most simplices belong to fractal structures.

Fig.~\ref{fig:sim2} shows the 
heights of consecutive 
simplices along loops with a growing sequence of winding numbers in 
the x-direction, starting from a simplex lying far within such a fractal structure,
called also an {\it outgrowth}.
We can see that as the winding number of the loop increases, 
usually the minimal x-height of the simplices belonging to the loop decreases,
ultimately down to 18, which is the length of the shortest
$\{1,0,0,0\}$ loops in the configuration.
As there are no two completely separate x-loops of length 18,
this means that all the loops of a high winding number pass many times mostly through the same set
of simplices.
The graphs showing the heights in the y, z and t directions for 
the same set of loops
demonstrate that although there is a correlation between height in all the directions,
the correlation is not perfect, especially after the loop leaves the outgrowth.
The repeating saw-like pattern is a loop of a high winding number tracing one of the shortest loops
of unit winding number several times. The heights of simplices belonging to a loop that is
shortest in the x-direction are not minimal in the other three directions.
However, even though in the other directions the height fluctuates,
it still remains close to the minimal height, 
as the simplices in the semi-classical region
have low heights in all directions.

\section{Alternative boundary conditions}\label{alternativeBC}

The other method, mentioned in Sec.~\ref{sec:higher}, is searching for a shortest path 
connecting a simplex to its copy in another elementary cell in a
system that is infinite in one direction and has periodic boundary
directions imposed in the other three directions.
The values of the winding numbers in the other directions are irrelevant for this method, i.e., we find loops with winding numbers of the form
$\{n,a,b,c\}$, $a,b,c$ being any integers, instead of only $\{n,0,0,0\}$. 

\begin{table}[H]
\begin{center}
\begin{tabular}{ |c|c|c|c|c|  }
\hline
\multicolumn{4}{|c|}{Winding numbers}&\multirow{2}{*}{Length}\\
 \cline{1-4}
 x&y &z&t&\\
 \hline
1   & 0    &0&   0 & 18\\
1 & -1 & 1 & 0 & 16\\
2 & -1 & 1&0 & 27\\ 
 \hline
 0 & 1 & 0 & 0 & 19 \\
 0 & 1 & 0 & 1 & 16 \\
 -1 & 1 & -1 & 0 & 16\\
 -2 & 3 & -1 & 1 & 43 \\
 \hline
 1 & 0& 0 & 0 & 18\\
 1 &-1& 1 &0 &16 \\
 2 & -1 & 3 & 0 & 47 \\
 \hline
\end{tabular}
\end{center}
\caption{Lengths of the shortest loops in the three basic spatial directions.}
\end{table}

It turns out that also for these boundary conditions paths starting at various simplices
tend to converge and follow a handful of very short loops. However, those loops are almost never
the shortest loops with unit winding numbers, e.g. $\{1,0,0,0\}$. Rather than that, those loops have
winding numbers of the form $\{n,a,b,c\}$ yet are shorter than n times the length of the 
shortest $\{1,0,0,0\}$ loop in the configuration (see Table 1; we used data from a diffusion in a
four-dimensional infinite system to find the precise winding numbers). One could conjecture
that as we probe loops of higher winding numbers we should eventually find
loops with even smaller ratio of total length to the winding number, but in fact
it turns out that even loops of winding numbers of order 50 utilize the loops
described in Table 1, so it appears that those loops are minimal.
It seems likely that the torus is twisted in such a way that it is impossible to
redefine the elementary cell so as to make loops in the basic directions always an optimal choice
as components for loops of higher winding numbers.

\begin{figure}[]

\centering

\includegraphics[width= 0.95 \columnwidth]{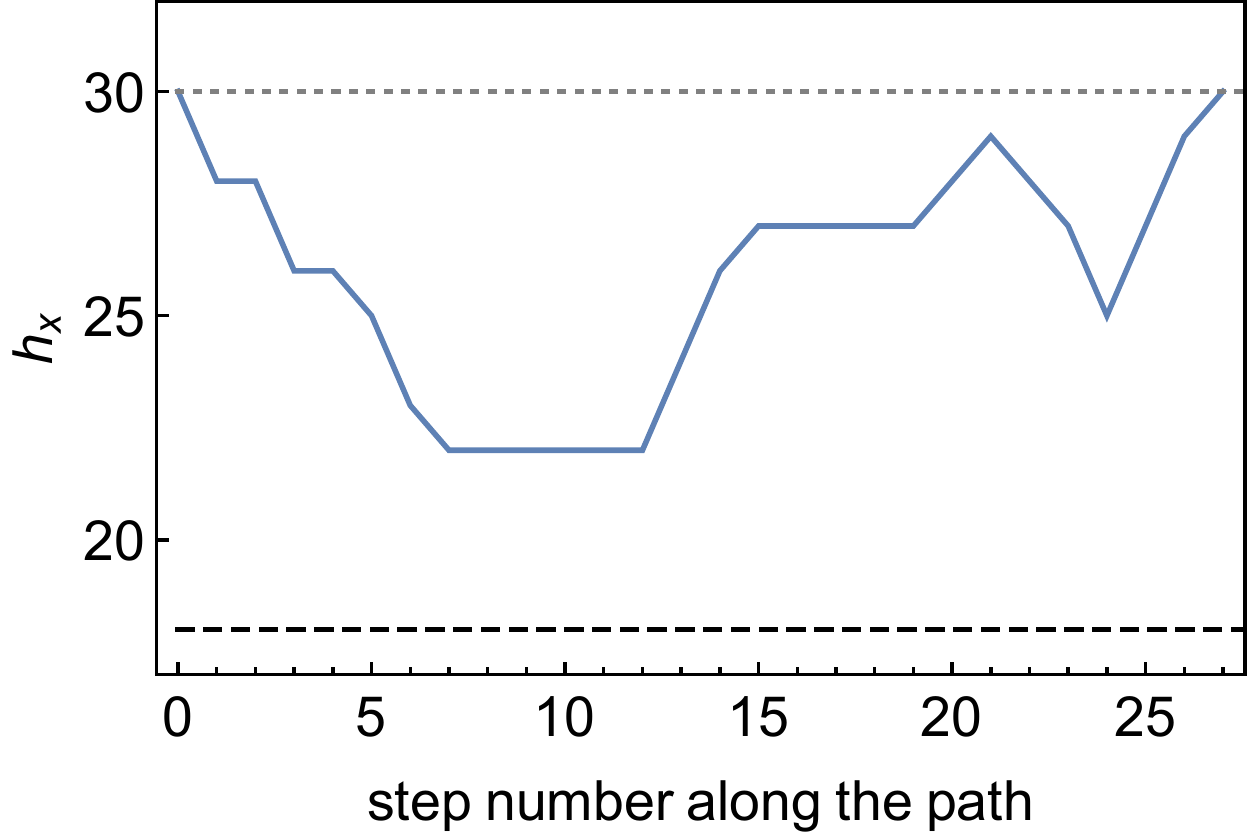} 
\includegraphics[width= 0.95 \columnwidth]{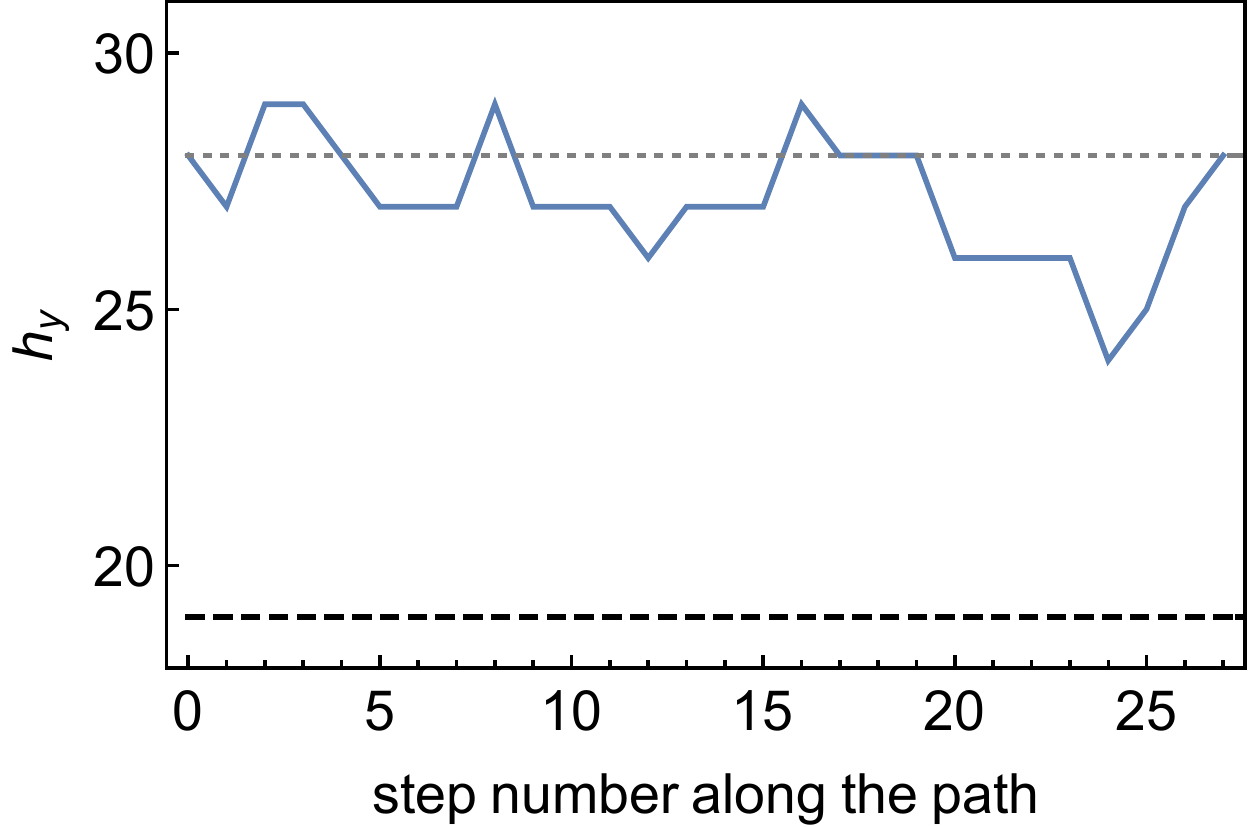}
\includegraphics[width= 0.95 \columnwidth]{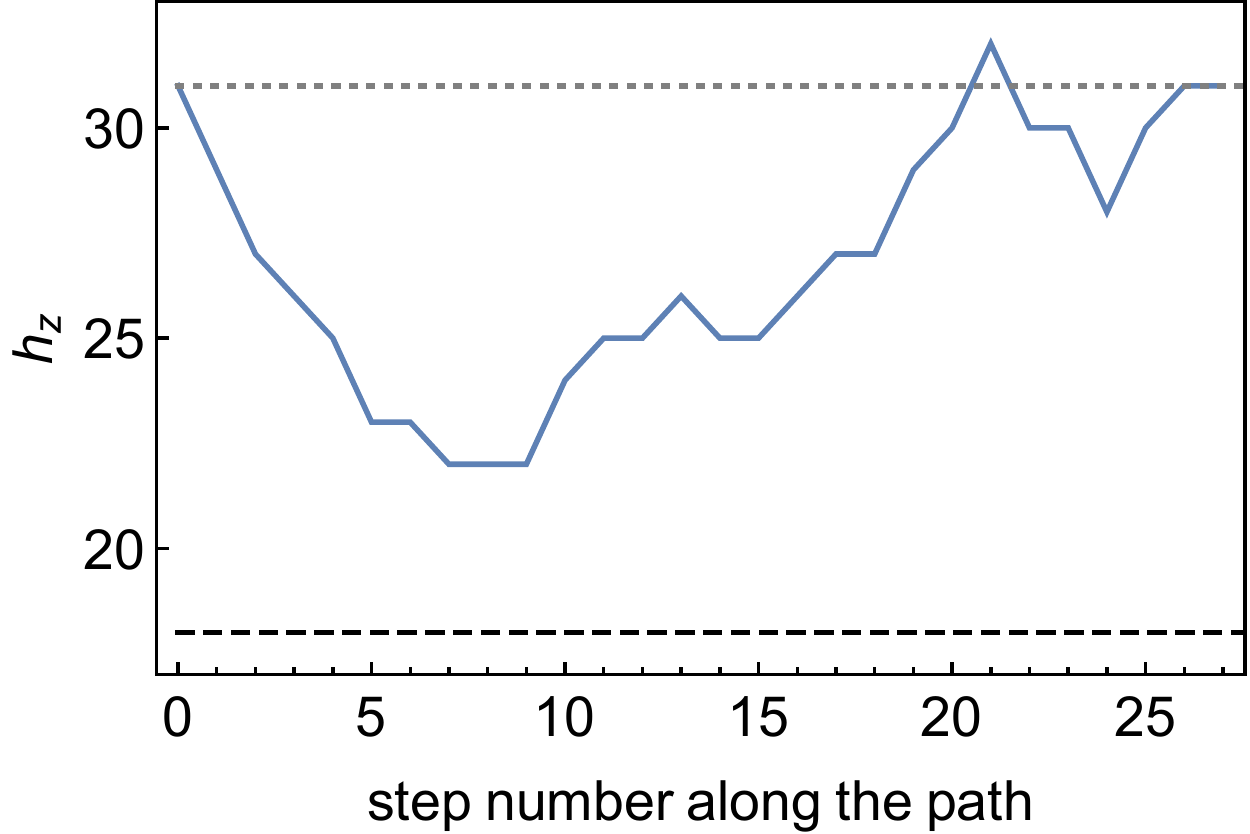} 
 \includegraphics[width=0.95 \columnwidth]{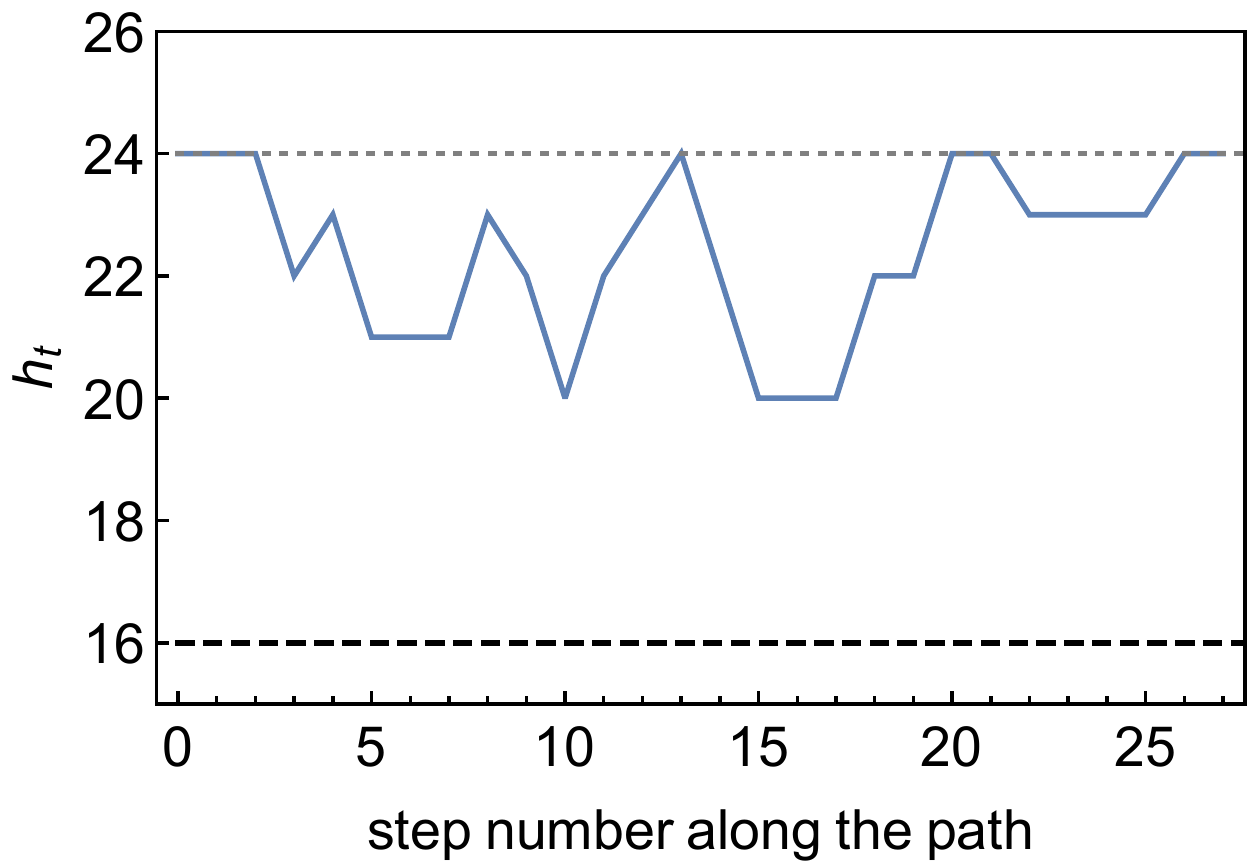}

\caption{The four basic heights of simplices belonging to 
one of the loops from Table 1: a loop of length 27 with winding numbers  $\{2,-1,1,0\}$.}
\label{fig:a}
\end{figure}

\begin{figure}[]

\includegraphics[width= 0.95 \columnwidth]{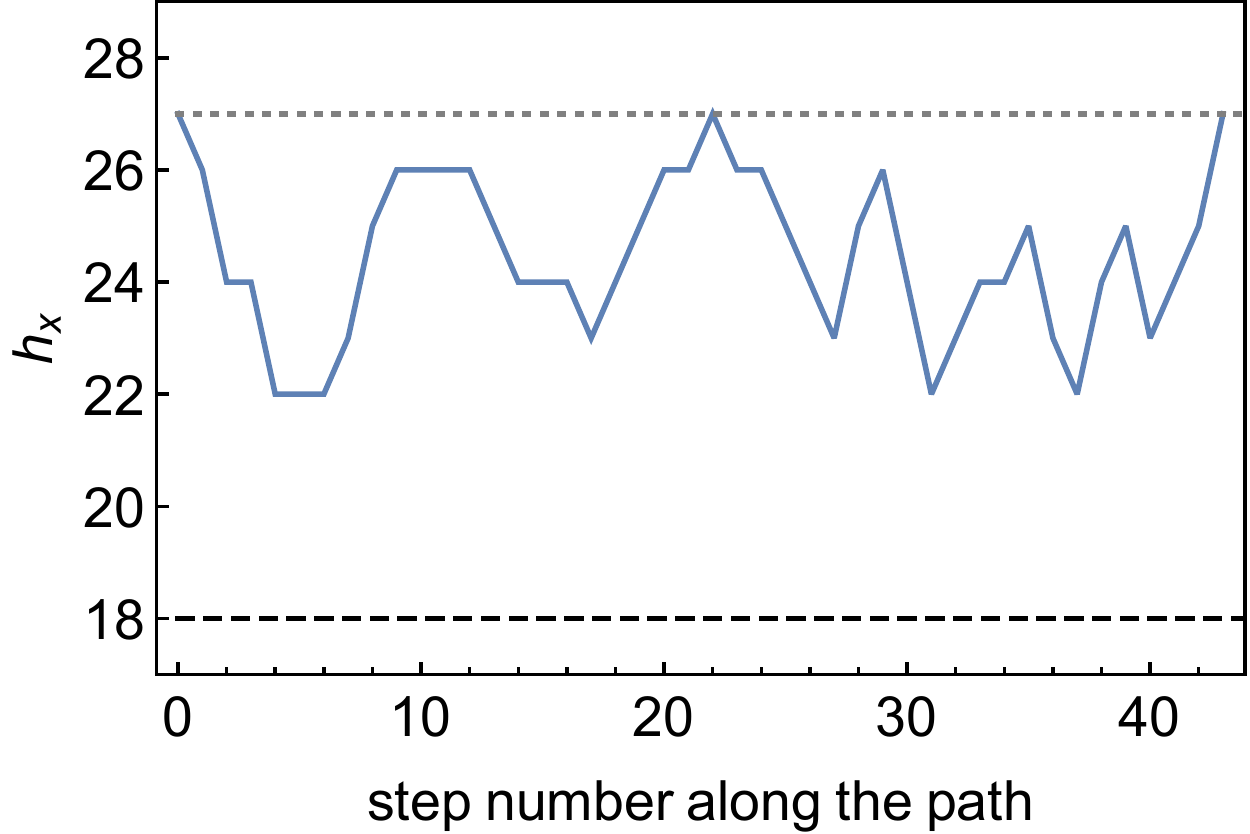} 
 \includegraphics[width= 0.95 \columnwidth]{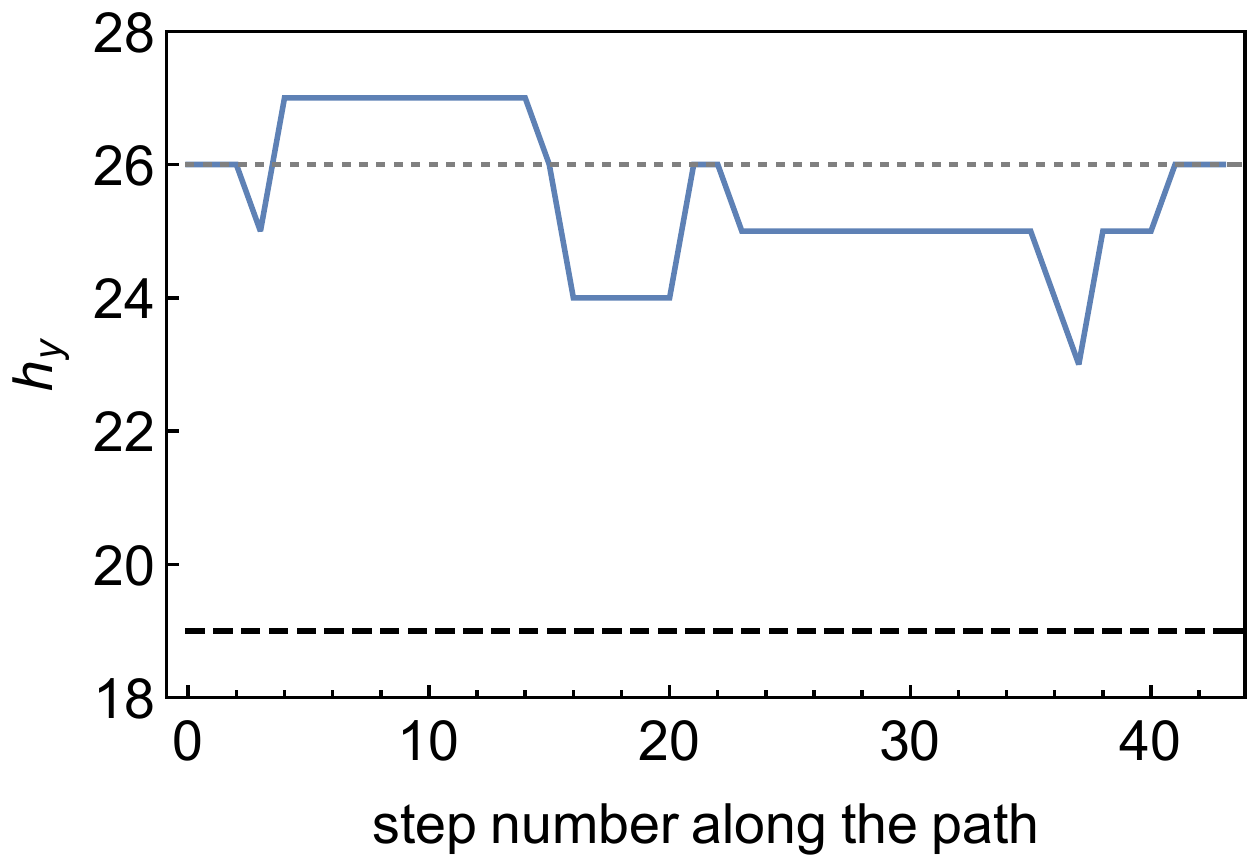}
\includegraphics[width= 0.95 \columnwidth]{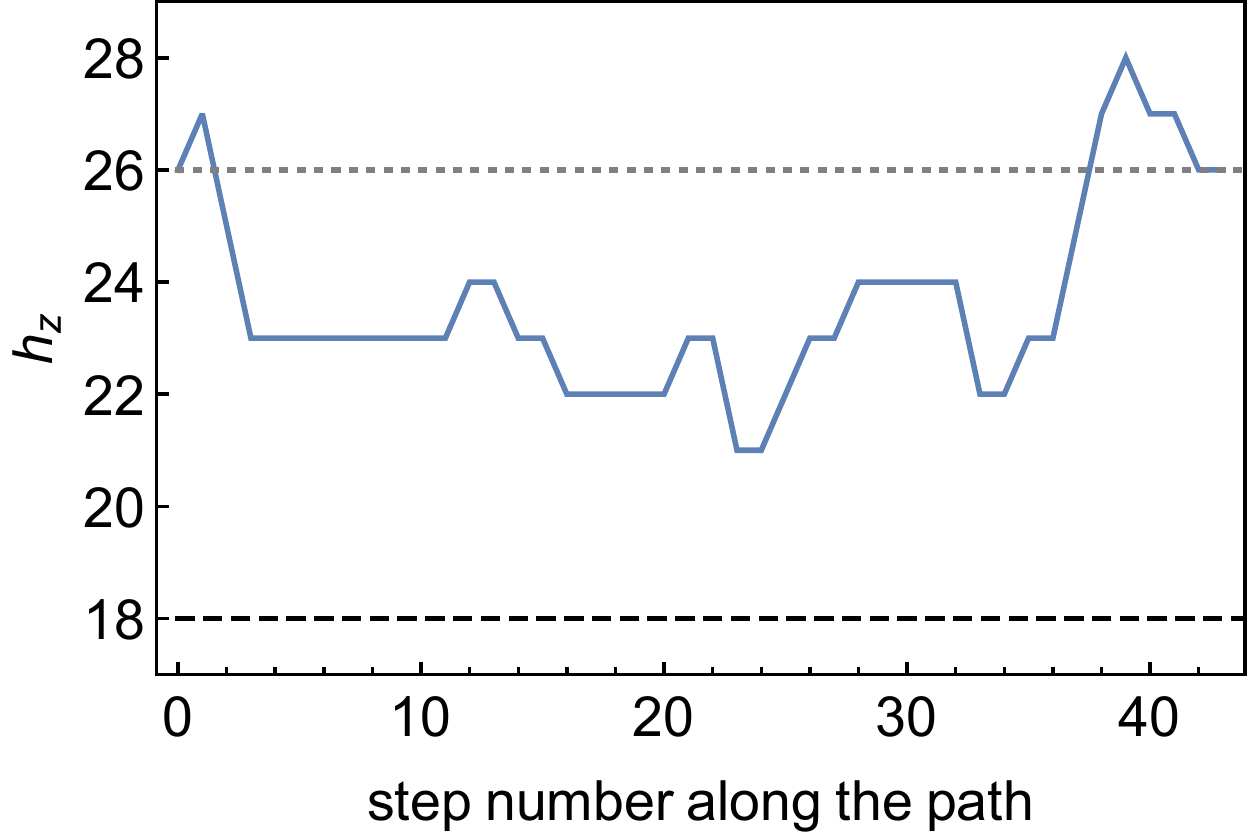} 
\includegraphics[width= 0.95 \columnwidth]{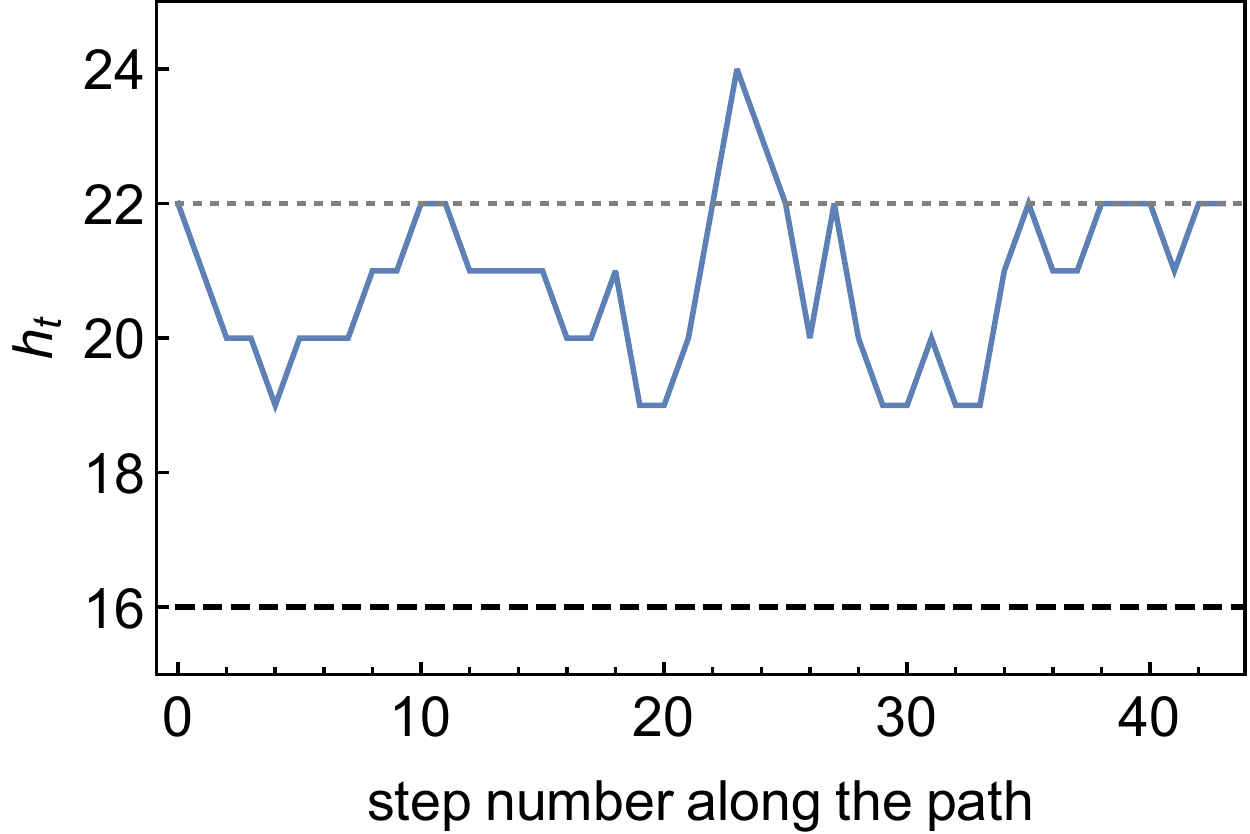}

\caption{The four basic heights of simplices belonging to one of the loops from Table 1: a loop of length 43 with winding numbers  $\{-2,3,-1,1\}$. }
\label{fig:b}
\end{figure}

\begin{figure}[]

\centering

\includegraphics[width= 0.95 \columnwidth]{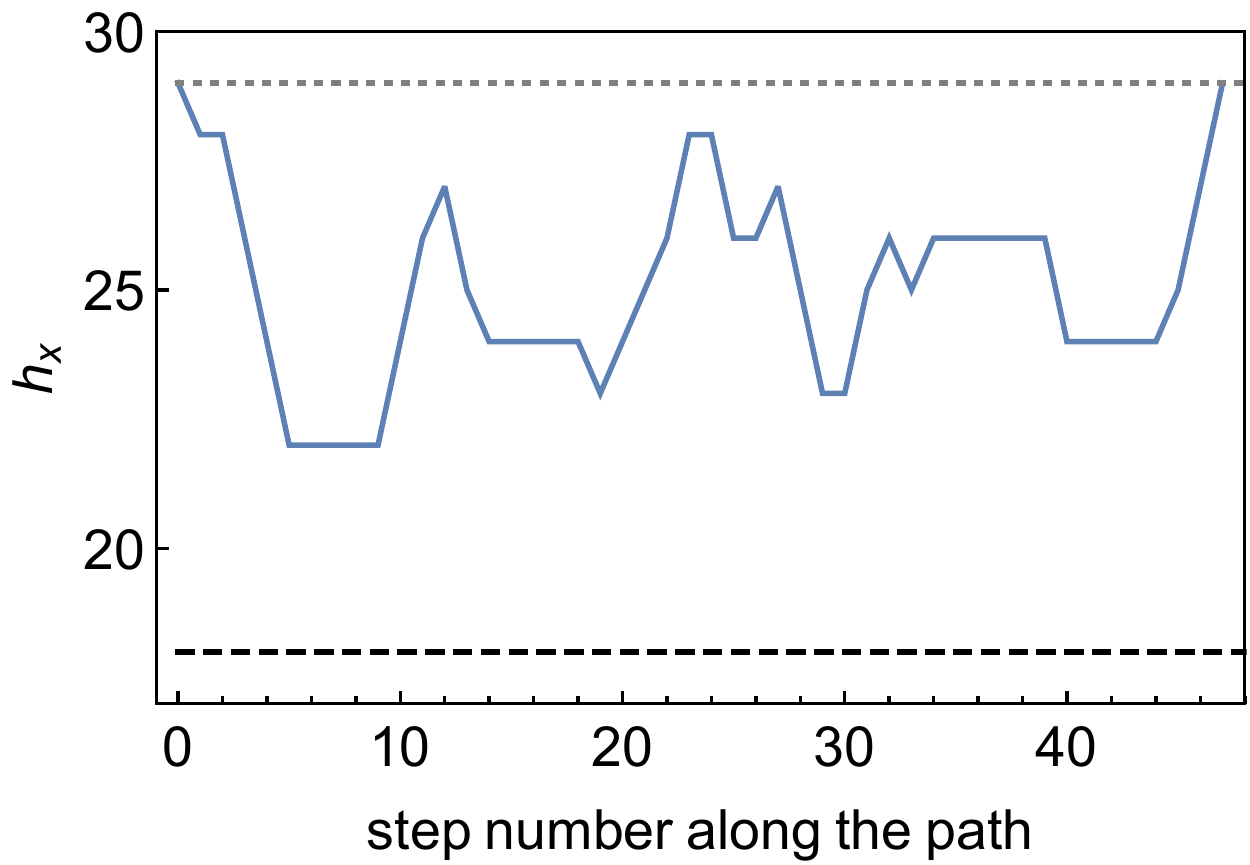} 
\includegraphics[width= 0.95 \columnwidth]{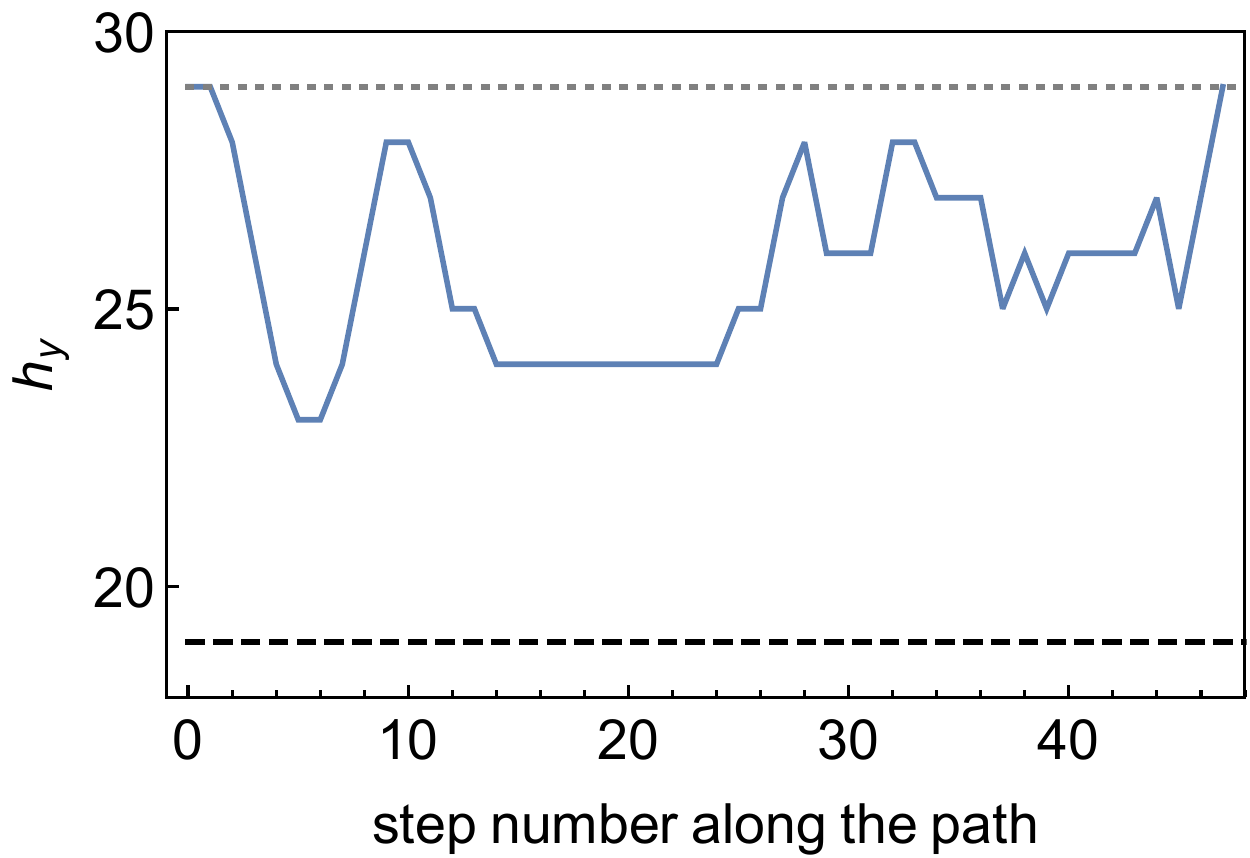}
\includegraphics[width= 0.95 \columnwidth]{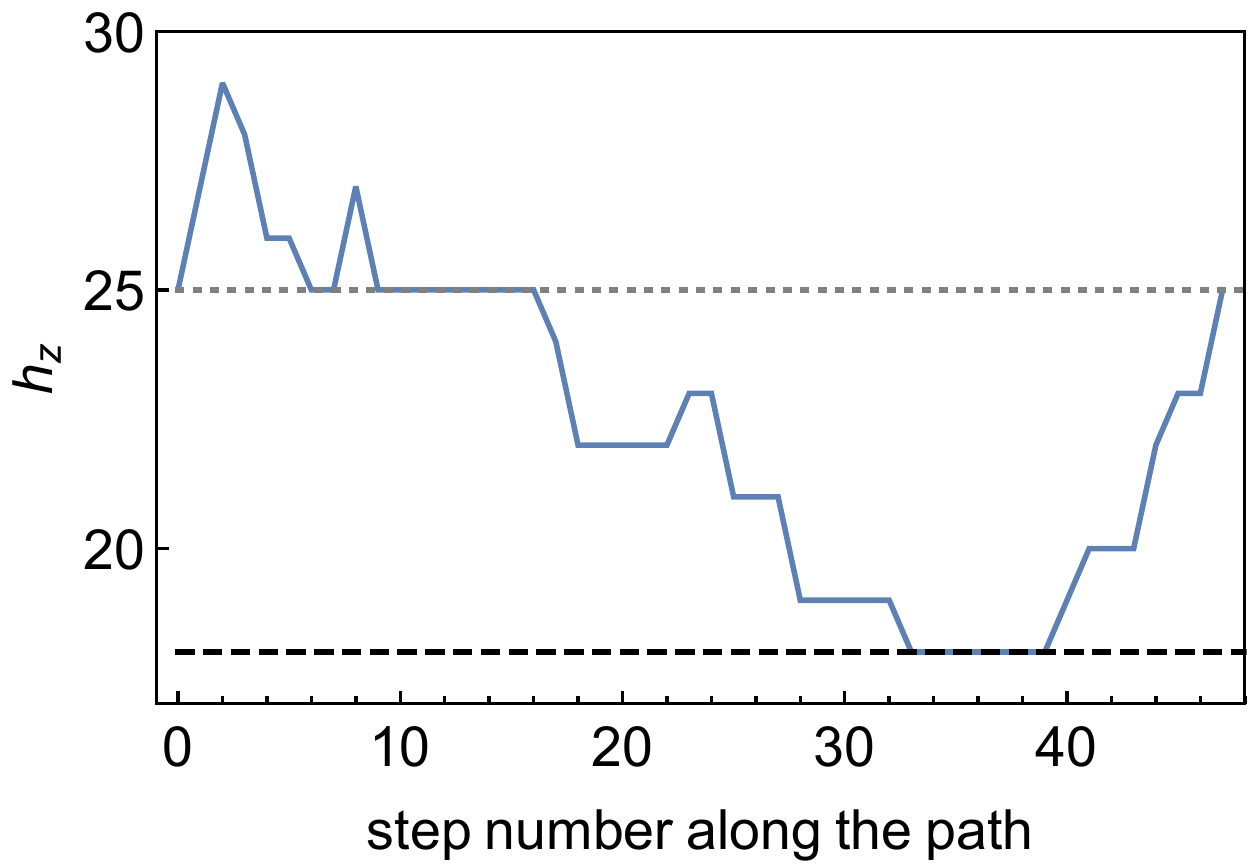} 
\includegraphics[width= 0.95 \columnwidth]{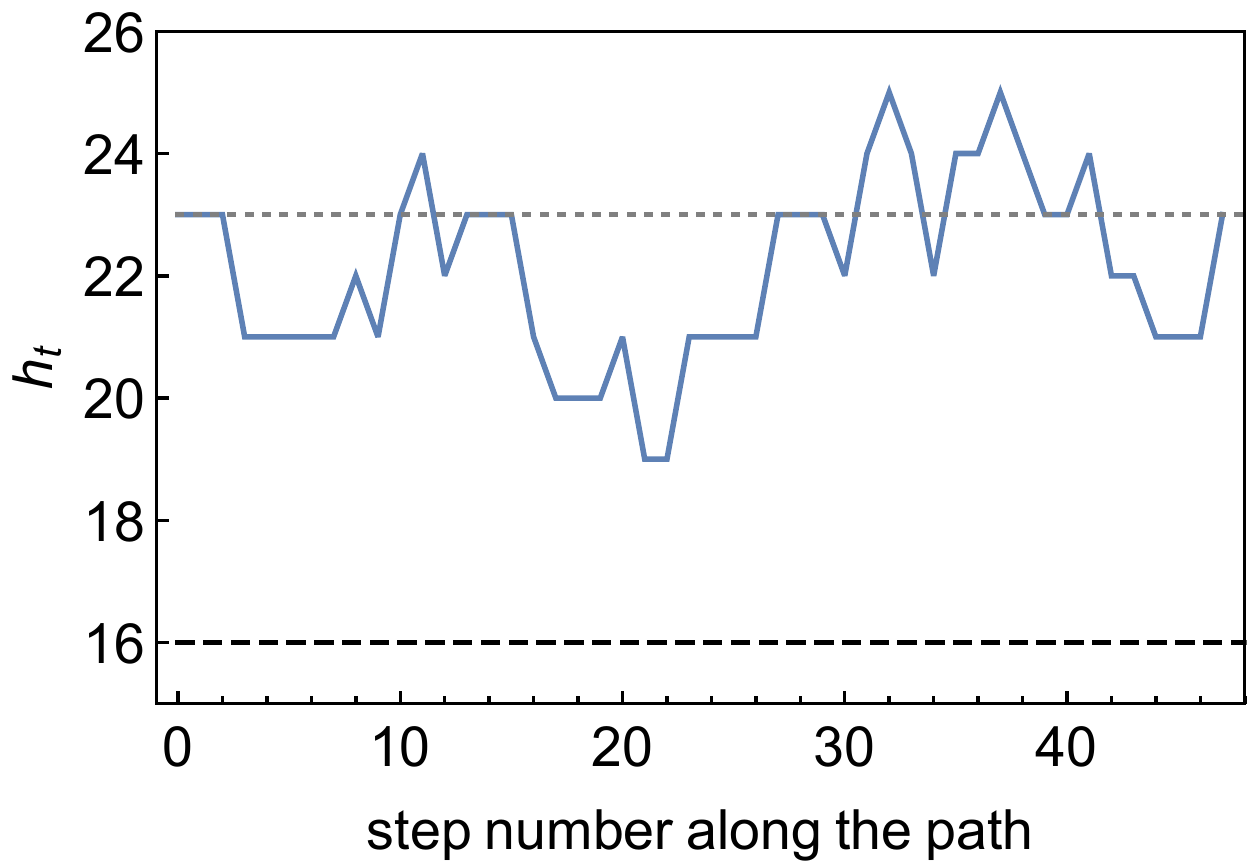}

\caption{The four basic heights of simplices belonging to one of the loops from Table 1: a loop of length 47 with winding numbers $\{2,-1,3,0\}$.}
\label{fig:c}
\end{figure}

Figs.~\ref{fig:a}-\ref{fig:c} show the heights in the basic directions of the consecutive simplices along
the shortest loops described above. We can see that the heights, while not minimal,
are quite low compared to the average (which is, as mentioned before, above 30). This signifies
that these paths too belong to the base (``bulk'') region of the torus
and are not composed of generic simplices, which mostly belong to the fractal structures.

\begin{figure}[]
\includegraphics[width= \columnwidth]{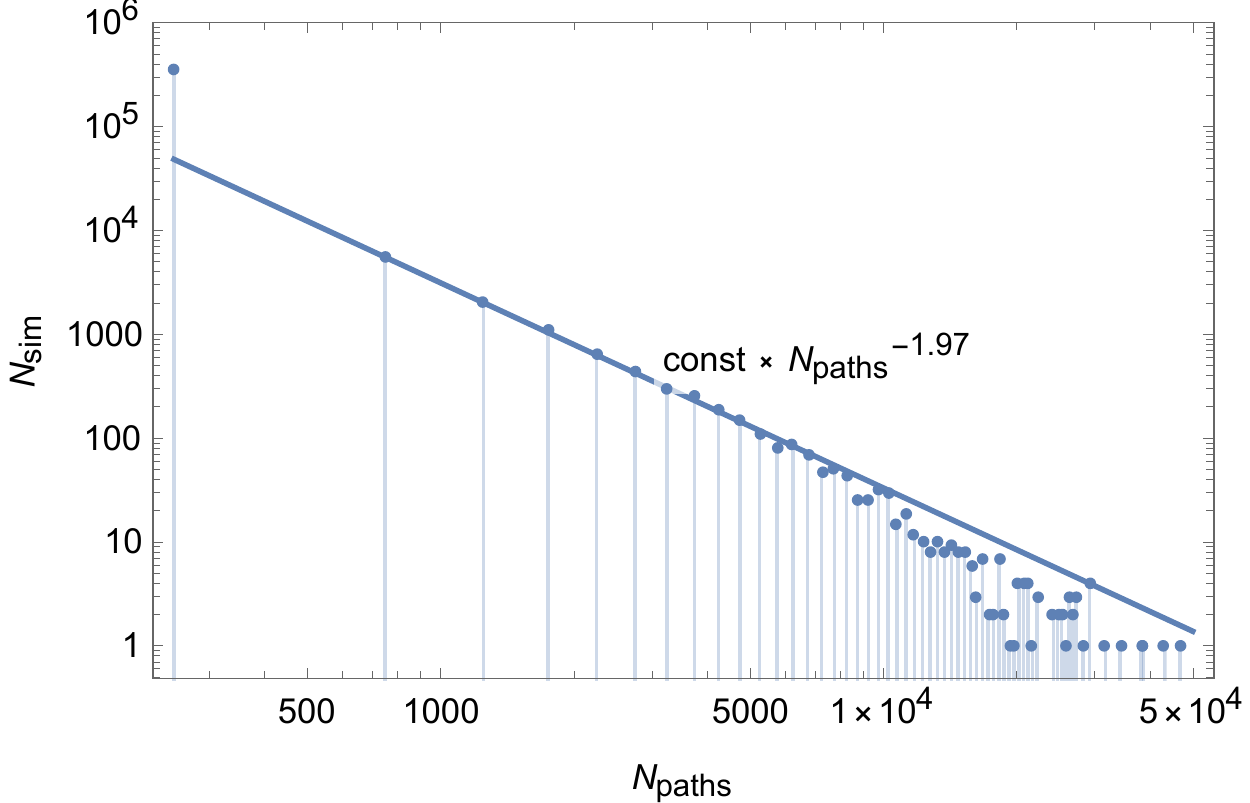}
\caption{A log-log scale histogram of the number of simplices
crossed by a given number of shortest loops.}
\label{fig:histogfit1}
\end{figure}

The algorithm we used creates a diffusion wave starting from a chosen simplex.
For each simplex reached in consecutive steps it stores one of the simplices from which it came.
In this way, we obtain at the same time not only the lengths of loops of winding number $n$ in the
chosen direction but also the lists of simplices along those loops. 
For each simplex in the configuration, we found and wrote down one shortest path connecting it with its nearest copy
in the x-direction. In this way we obtained the lists of simplices
belonging to 370724 shortest loops. We repeated the same process in the y- and z-directions.
Next, we removed from each list the initial and final simplex of each loop, and we counted
the number of appearances of each simplex in the lists. 
A log-log scale histogram is shown in Fig.~\ref{fig:histogfit1}.
The maximum value was more than 40000, which 
corresponded to one of the bulk simplices,
and the minimal value was zero, which occurred numerous times and corresponded to simplices at the far ends
of the outgrowths. The latter simplices are not a part of any geodesics apart from those that start within them.
We were able to fit to the histogram a power law curve with the exponent very close to $-2$, which seems to bear a certain 
significance. This functional relationship is different than in the cases of, e.g., a branched polymer or a regular lattice.
We have not yet found an algorithmic method of constructing a toroidal graph with behavior described 
by the same exponent. We plan to investigate 
this point in a future article.

\begin{figure}[]
\includegraphics[width= \columnwidth]{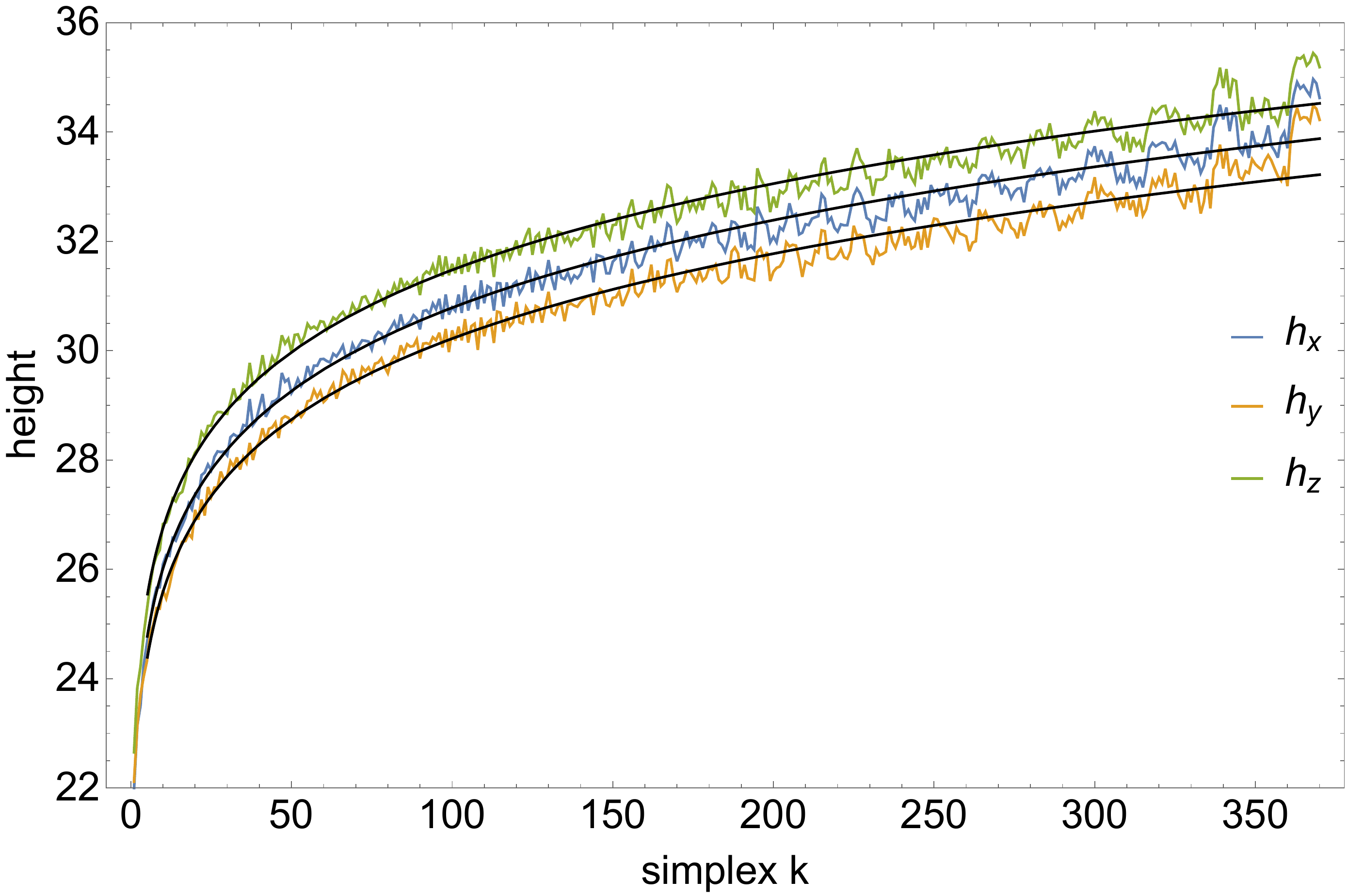}
\caption{The x-, y-, and z-heights of simplices. The simplices were sorted in the order of
descending number of loops passing through them, and then the heights were averaged over blocks
containing 1000 simplices each. The fitted functions are $h_x=21.98 k^{0.0732}$, $h_y=21.66 k^{0.0723}$ and $h_z=22.74 k^{0.0706}$.}
\label{fig:histogfit2}
\end{figure}

We also noted that the heights of the simplices and the number of shortest loops passing through them
are strongly correlated. We sorted the simplices in the order of descending number of loops passing through them, 
divided the list into blocks containing 1000 simplices each, and within each block
averaged the heights in each of the three spatial directions. With this ordering of simplices, the heights
turned out to be increasing functions of the ordinal number of simplices in the list, modulo statistical
fluctuations (see Fig.~\ref{fig:histogfit2}). The fluctuations in all three directions were strongly correlated.
We fitted a power law to the curves. The exponent is probably the same for all the directions, 
and the constant factor depends on the shape of the torus – it is higher for the directions in which
the torus is more elongated (and so the average height of the simplices is greater).

This shows that the number of loops passing through a given simplex can serve as another indicator
of its position within the torus. Most of the geodesics between distant points
pass through the bulk simplices in the semi-classical region
and do not enter the outgrowths, which are the regions of quantum fluctuations.
If a geodesic passes through a simplex in the outgrowth, it usually means that it had its beginning even
deeper in the same outgrowth.

\section{Neighborhoods of bulk- and outgrowth-simplices}\label{neighborhood}

As observed, geodesics between distant simplices tend to pass through simplices of low heights
and avoid simplices of middle and greater heights. It is tempting
to interpret the former as belonging to a semi-classical ``bulk'',
and the latter to ``outgrowths'', which are a result of quantum fluctuation.
If so, then those regions should differ also in other properties apart from such 
non-local ones as height and
the number of loops passing through the simplices.
This is in fact the case.
The neighborhoods of bulk simplices and outgrowth simplices 
look considerably different, which allows for a construction
of local quantities distinguishing between them.

\begin{figure}[H]
\includegraphics[width= \textwidth]{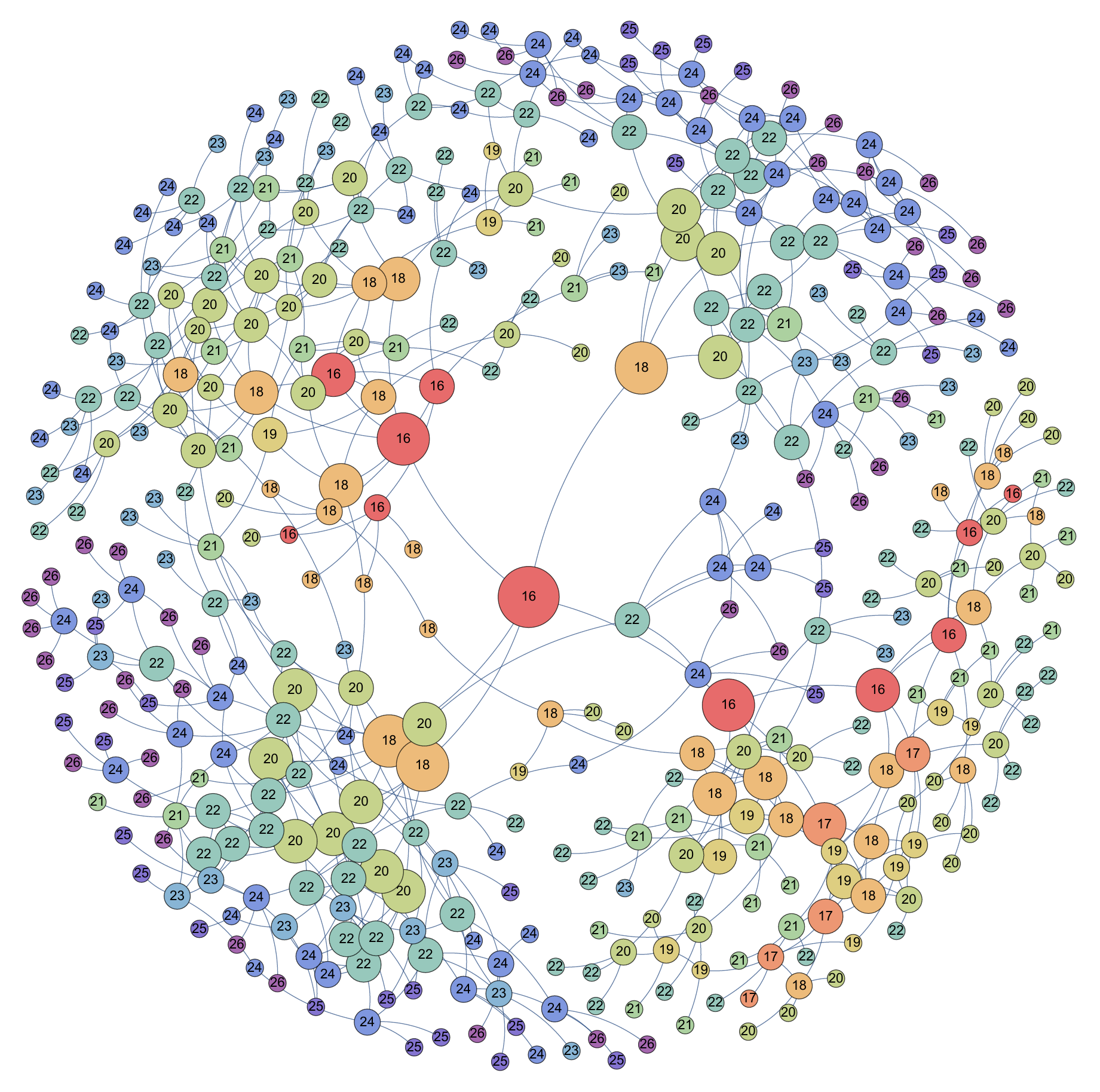}
\caption{Six concentric shells around a simplex of height equal to 16.  The colors and the numbers indicate the height of a simplex, and the size of a vertex its distance from the starting simplex. The central simplex lies in the bulk region.}
\label{fig:83621}
\end{figure}

\begin{figure}[]
\includegraphics[width= \textwidth]{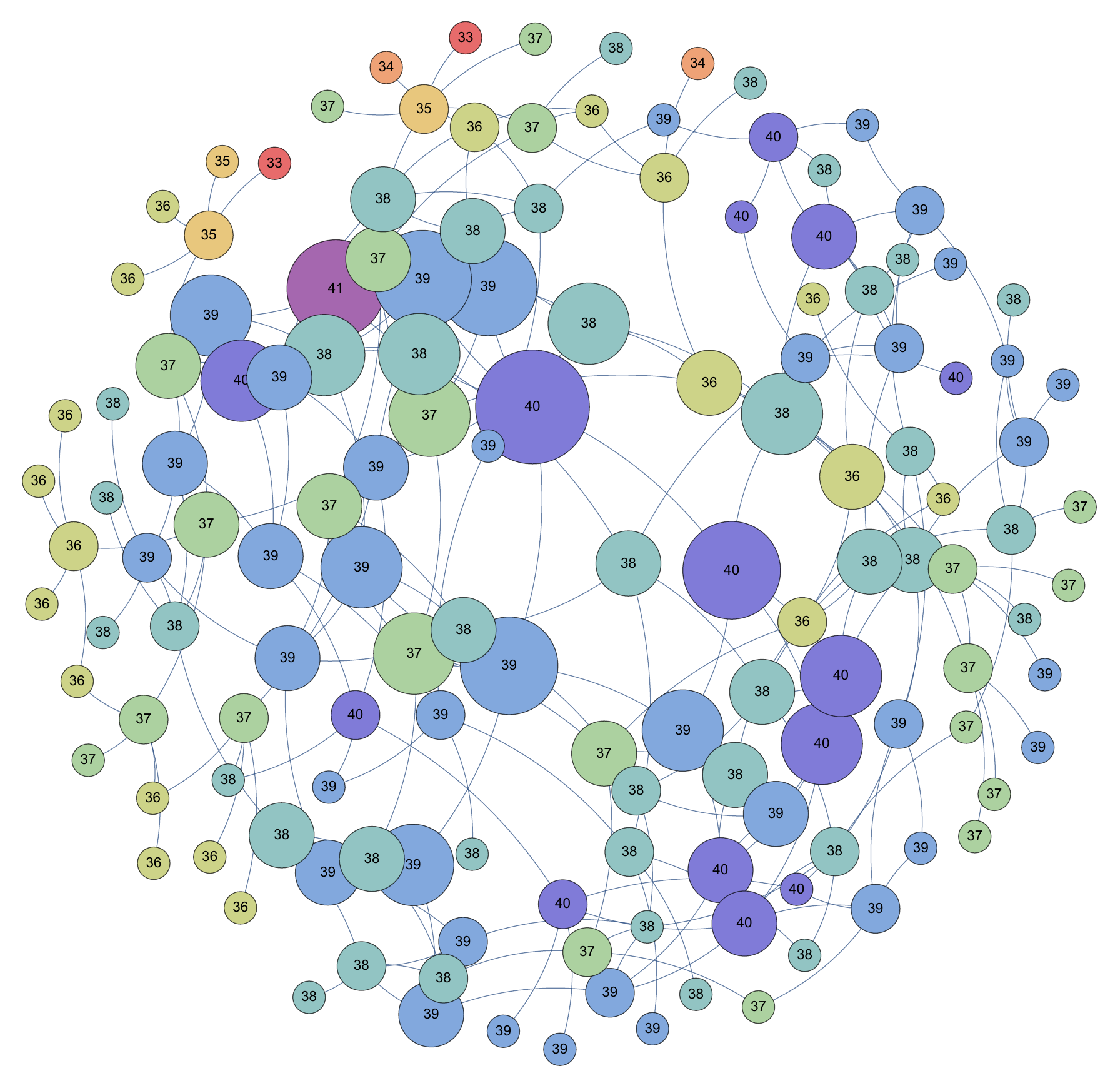}
\caption{Six concentric shells around a simplex of height equal to 40.  The colors and the numbers indicate the height of a simplex, and the size of a vertex its distance from the starting simplex. The central simplex lies near the deep end of an outgrowth.}
\label{fig:113}
\end{figure}

Figs.~\ref{fig:83621} - \ref{fig:113} show subgraphs of the dual lattice, depicting all the simplices
of distance up to 6 from a starting simplex, 
together with the connections between them.
The simplices – vertices of the dual lattice – are represented by circles whose size is a decreasing 
function of distance from the starting simplex. Colors of the circles indicate heights, with red corresponding to the shortest and violet to the longest loops in a given figure. 
The heights are also noted as numbers in the circles.\footnote{
Figures in this section are based on data obtained using the version of algorithm
with periodic boundary conditions in directions y, z and t. 
Heights defined in this alternative way have similar values and interpretation to x-heights
defined previously, though they are not directly equivalent,
e.g., here the minimal value is 16,
which occurs in simplices that form the shortest loop with winding numbers \{1,-1,1,0\} (see Table 1).}

In Fig.~\ref{fig:83621} the starting simplex had
height equal to 16. 
As that is the minimal value, we can see that in the neighborhood of the simplex height tends
gradually to increase together
with distance from it. By the same token, loop length decreases with increasing distance from a simplex
with height equal to 40, which is among largest in the configuration – Fig.~\ref{fig:113}. 
Even a cursory glance at the graphs, moreover, suffices to note
a strong dependence of the total number of simplices
of distance up to 6 from the starting simplex on its loop-length.
As expected, ``outgrowth'' regions have an elongated shape and 
a lower Hausdorff dimension than the ``bulk'' region, which
is a sign of their fractality.

\begin{figure}[]
\includegraphics[width= \textwidth]{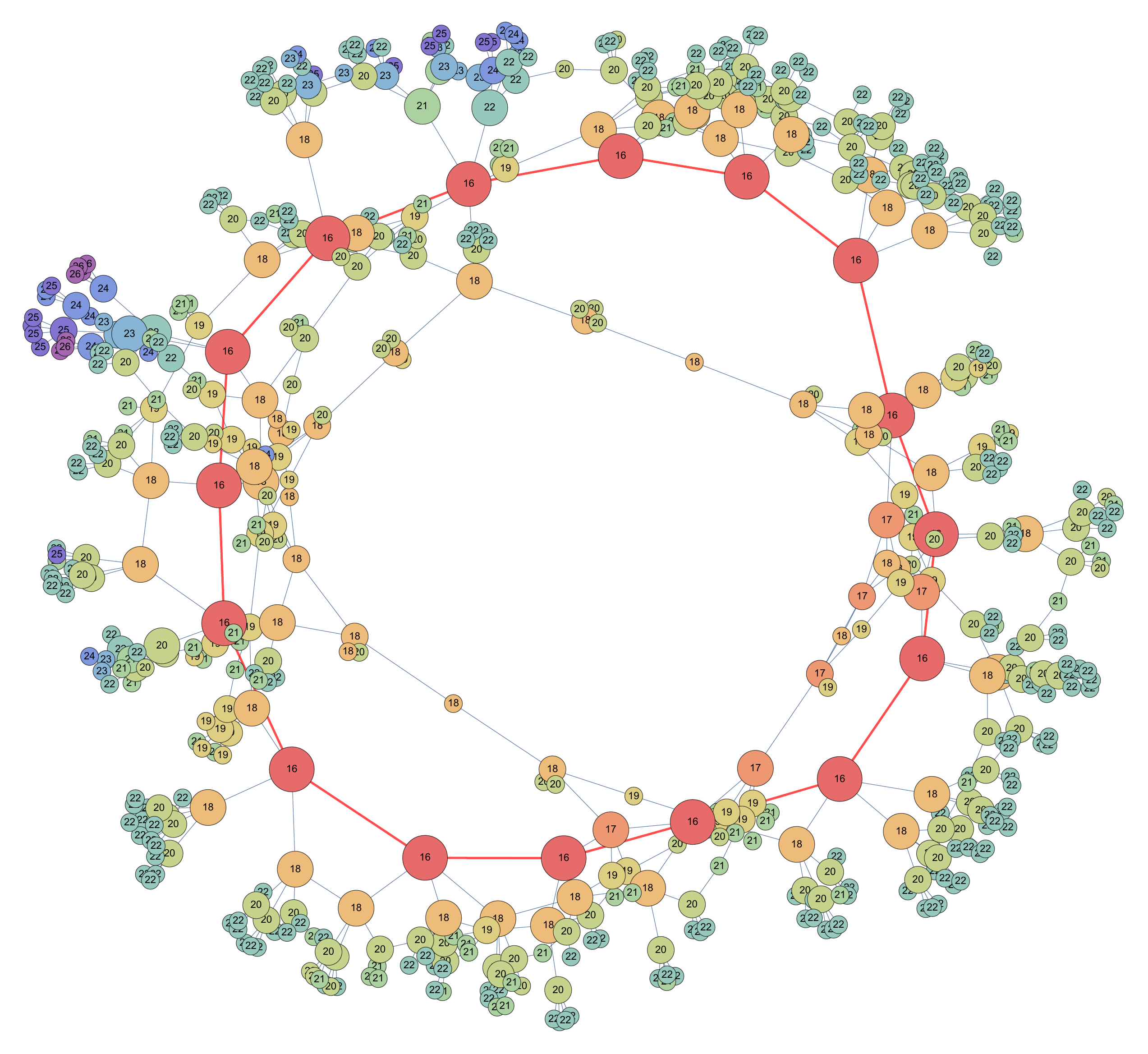}
\caption{The shortest \{1,-1,1,0\} loop together with its neighborhood. The colors and the numbers indicate the height of a simplex, and the size of a vertex its distance from the loop.}
\label{fig:loopgraph}
\end{figure}

Fig.~\ref{fig:loopgraph} shows the shortest \{1,-1,1,0\} loop and its neighborhood. Comparing it
also with Table 1, we note that it is
among the very shortest non-trivial loops in the configuration.
It is the only loop with that set of winding numbers and length 16. It is readily seen
that as we count loops with the same winding numbers and length 17, 18, etc.~in 
the vicinity of the marked loop, the number grows approximately exponentially.

\onecolumngrid

\twocolumngrid

\onecolumngrid

\twocolumngrid
\onecolumngrid

\twocolumngrid

\onecolumngrid

\begin{figure}[]
\includegraphics[width= \textwidth]{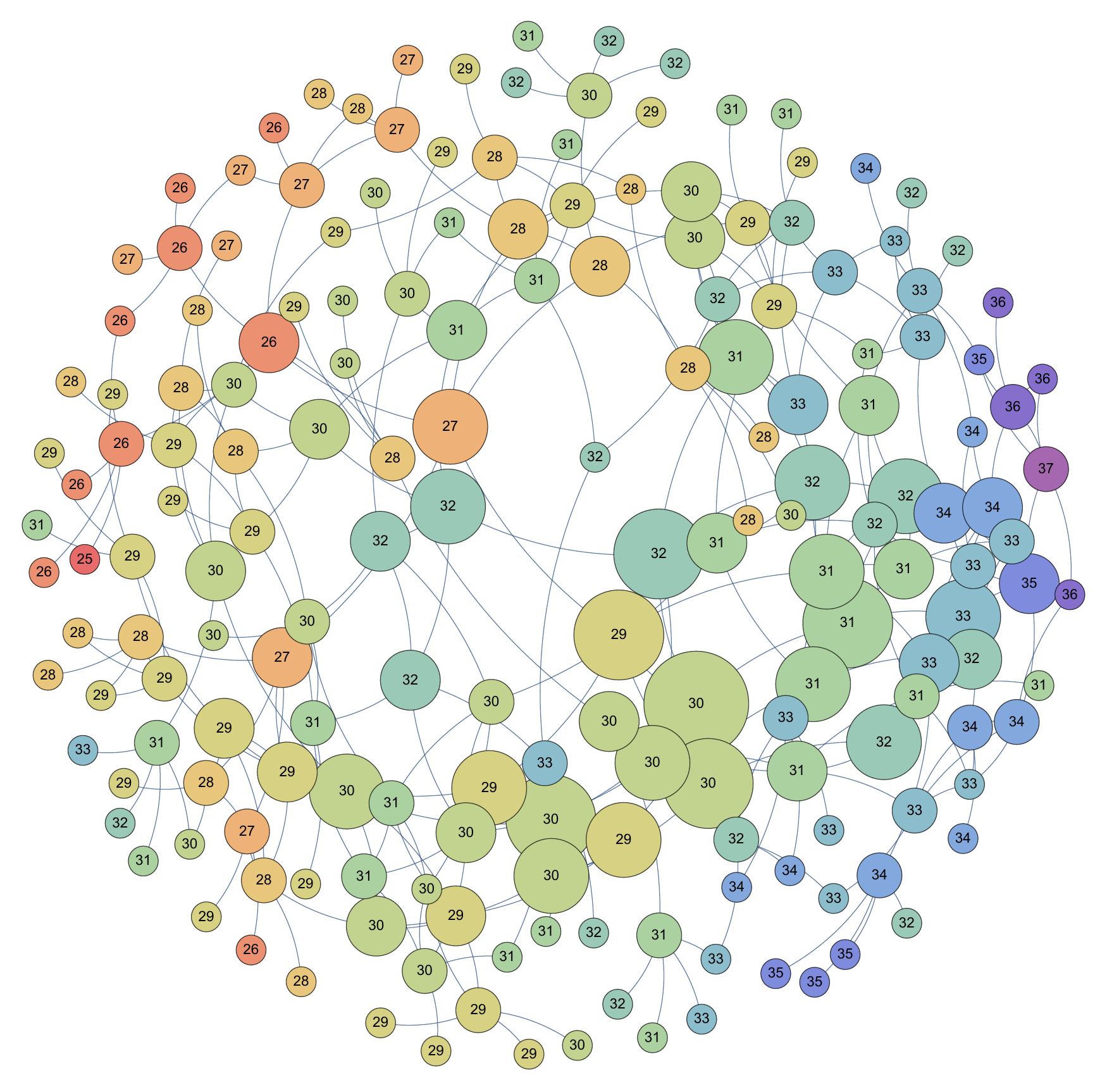}
\caption{Six concentric shells around a simplex of height equal to 30.  The colors and the numbers indicate the height of a simplex, and the size of a vertex its distance from the starting simplex. The central simplex lies in the middle of an outgrowth.}
\label{fig:25}
\end{figure}

\twocolumngrid

\section{Conclusion}\label{conclusion}

The detailed measurements performed on a typical toroidal configuration which appears in the CDT path integral in the $C$ phase shows the following: the spatial $T^3$ part 
consists of a relatively small 
bulk region, {\it the toroidal center}, which we have denoted semi-classical, and numerous
fractal outgrowths of almost spherical topology (with a single small boundary)
which contains 
most of the simplices. A lower-dimensional illustration of this 
is shown in the upper part of Fig.~\ref{fignew}. Introducing
the lengths of the shortest non-contractible loops in 
the coordinate directions as the heights associate with 
a given simplex allowed us to classify the simplex as belonging
to the toroidal center or to an outgrowth. Further, the number 
of simplices in the outgrowths where the height is a local 
maximum is not small. The interpretation of this is that the 
outgrowths are quite fractal, again in the way illustrated in 
the upper part of  Fig.~\ref{fignew}. An important feature
of the length distributions of non-contractible loops associated
to the simplices is that they scale as $N^{1/4}$, where 
$N$ denotes the size of the triangulation, i.e., the number of 
four-simplices, as shown in Fig.~\ref{fig:shifts}. The most likely
consequence of such a ``canonical'' scaling is that the
volume of the toroidal center, although small compared to the
volume of the outgrowths,  will also scale with $N$. It might 
thus be justified to think of it as semi-classical, in 
contrast to the toroidal center-part of the two-dimensional 
configuration shown in the lower part of  Fig.~\ref{fignew},
which vanishes in the large $N$ limit, as discussed in the Introduction.

We conclude that there is a well defined geometric structure
underlying the typical path integral configuration of $T^4$ in 
CDT. It is somewhat more fractal than we had hoped for in the sense that the outgrowths contain most of the simplices, but it 
invites to use a classical scalar field to define a coordinate system, a procedure common in classical General Relativity. By imposing suitable boundary conditions on the scalar field 
one can make its values record the structure of the 
toroidal center well, whereas the field is almost constant in 
an outgrowth. It thus emphasizes what we have denoted 
the semi-classical part of the configuration and might be 
a good choice of coordinates if one wants to construct a 
semi-classical action. Work in this direction will be reported
elsewhere.

\section*{Acknowledgement}
The authors thank Jakub Gizbert-Studnicki and Daniel N\'emeth for many fruitful discussions.
JA  acknowledges support from the Danish Research Council grant {\it Quantum Geometry},
grant 7014-00066B. ZD acknowledges support from the National Science Centre, Poland, grant 
2019/32/T/ST2/00390.
AG acknowledges support by the National Science Centre, Poland, under grant no. 2015/17/D/ST2/03479.
JJ acknowledges support from the National Science Centre, Poland,
grant  2019/33/B/ST2/00589.

\bibliography{winding}

\end{document}